\documentclass{article}

\usepackage{arxiv}

\usepackage[utf8]{inputenc} 
\usepackage[T1]{fontenc}    
\usepackage{hyperref}       
\usepackage{url}            
\usepackage{booktabs}       
\usepackage{amsfonts}       
\usepackage{nicefrac}       
\usepackage{microtype}      
\usepackage{lipsum}		
\usepackage{graphicx}
\usepackage[numbers,sort&compress]{natbib}
\usepackage{doi}
\usepackage{subcaption}
\usepackage{amsmath}
\usepackage{amssymb}
\usepackage{nameref}
\usepackage{algorithm}
\usepackage{algpseudocode}
\usepackage{comment}
\usepackage{tikz}
\usepackage{mathtools}
\usepackage{setspace}
\usepackage[toc,page]{appendix}

\title{Energy-Based Coarse-Graining in Molecular Dynamics: A Flow-Based Framework without Data}

\author{ \href{https://orcid.org/0000-0002-1657-3060}{\includegraphics[scale=0.06]{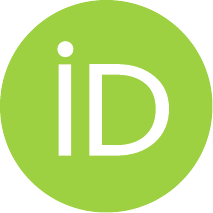}\hspace{1mm}Maximilian ~Stupp}\thanks{ Boltzmannstr. 15, 85748 Garching b. München, Germany} \\
	Professorship of Data-driven Materials Modeling\\
    School of Engineering and Design\\
    Technical University of Munich\\
	\texttt{maximilian.stupp@tum.de} \\
	\And
	\href{https://orcid.org/0000-0002-9345-759X}{\includegraphics[scale=0.06]{orcid.pdf}\hspace{1mm}P. S.~Koutsourelakis}\footnotemark[1] \\
	Professorship of Data-driven Materials Modeling\\
    School of Engineering and Design\\
    Munich Data Science Institute (MDSI - Core Member)\\
    Technical University of Munich\\
	\texttt{p.s.koutsourelakis@tum.de}
}

\date{}

\newcommand{\bs}{\boldsymbol}
\newcommand\bz{\mathbf{z}}
\newcommand\bx{\mathbf{x}}
\newcommand\bX{\mathbf{X}}
\newcommand{\rfeq}[1]{Equation \eqref{#1}}
\newcommand\bt{\bs{\theta}}
\newcommand\be{\begin{equation}}
\newcommand\ee{\end{equation}}
\newcommand\bphi{\boldsymbol{\phi}}
\newcommand\bepsilon{\boldsymbol{\epsilon}}
\newcommand{\pa}{\partial}
\newcommand\numberthis{\addtocounter{equation}{1}\tag{\theequation}}

\usepackage{color}

\hypersetup{
pdftitle={Energy-Based Coarse-Graining in Molecular Dynamics},
pdfsubject={physics.chem-ph},
pdfauthor={Maximilian ~Stupp, P. S.~Koutsourelakis},
pdfkeywords={Coarse-graining, Boltzmann distribution, Energy training, Normalizing Flow, Tempering},
}

\begin{document}
\maketitle

\begin{abstract}
Coarse-grained (CG) models provide an effective route to reducing the complexity of molecular simulations, but conventional approaches depend heavily on long all-atom molecular dynamics trajectories to adequately sample configurational space. This data dependence limits accuracy and generalizability, as unvisited configurations remain excluded from the resulting CG models.

We introduce a fully data-free, generative framework for coarse-graining that directly targets the all-atom Boltzmann distribution. The model defines a structured latent space comprising \emph{slow} collective variables, associated with multimodal marginal densities capturing metastable states, and \emph{fast} variables, represented through simple, unimodal conditional distributions. A learnable, bijective map from latent space to atomistic coordinates enables the automatic and accurate reconstruction of molecular structures. Training relies solely on the interatomic potential and minimizes the reverse Kullback–Leibler (KL) divergence via an energy-based objective. To stabilize optimization and ensure mode coverage, we employ an adaptive tempering scheme that promotes the exploration of diverse configurations. Once trained, the model can generate independent, one-shot equilibrium samples at full atomic resolution.

Validation on two synthetic systems, a double-well potential and a Gaussian mixture model, as well as on the benchmark alanine dipeptide, demonstrates that the method captures all relevant modes of the Boltzmann distribution, reconstructs atomic configurations with high fidelity, and automatically learns physically meaningful CG representations. These results suggest that the proposed framework provides a promising, data-free alternative to traditional CG techniques, offering both a principled approach to addressing the long-standing ``chicken-and-egg'' challenge in coarse-graining and an effective solution to the back-mapping problem by enabling accurate reconstruction of all-atom configurations.
\end{abstract}


\section{Introduction}
\label{sec:introduction}
The ability to predict molecular properties from first principles relies on our capacity to sample Boltzmann-weighted ensembles accurately. Molecular dynamics (MD) and Monte Carlo (MC) simulations provide frameworks for such sampling, offering a means to explore the thermodynamic and kinetic landscapes of complex biophysical systems \cite{alder_md_2004, metropolis_monte_mcmethod_1949}. Yet, as system complexity grows---such as in the case of drug-protein interactions or enzymatic catalysis---the time scales required to observe biologically relevant events far exceed what is accessible by brute-force simulations.
To overcome these limitations, coarse-graining (CG) has emerged as a crucial methodology that simplifies molecular representations by reducing the number of degrees of freedom (DOF), allowing for more efficient simulations while preserving key physical properties \cite{voth_coarse-graining_2008}. There are two main approaches: top-down and bottom-up methods \cite{noid_perspective_cg_pot_2013,jin_bottom-up_2022}. Top-down methods design CG models to reproduce specific macroscopic properties based on experimental data. In contrast, bottom-up coarse-graining techniques derive CG interactions by defining a mapping from the all-atom, fine-grained (FG) representation to a reduced, coarse-grained description \cite{kalligiannaki_geometry_2015}. Typically, this involves lumping multiple atoms into a pseudomolecule, often referred to as ``beads''. This inevitably results in a loss of information between the two scales \cite{katsoulakis_information_2006, foley_impact_2015} and makes recovering the all-atom structures from the CG representation a challenging back-mapping problem \cite{chen_automatic_2006, lombardi_cg2aa_2016, machado_sirah_2016, schoebel_predictive_CG_2017, peng_backmapping_2019, wang_generative_2022, jones_diamondback_2023}.

The second  necessary component in  bottom-up methods is defining a  model for the CG coordinates, which should reproduce the equilibrium distribution of the CG DOFs, known as thermodynamic consistency \cite{noid_multiscale_2008}. Many classical CG methods achieve consistency by finding an approximation of the potential of mean force (PMF) or the gradients thereof, i.e., the forces between the CG beads. Direct and Iterative Boltzmann Inversion \cite{Tschöp199861, IBI_2003} and Inverse Monte Carlo \cite{Lyubartsev19953730} are commonly used to derive effective CG potentials that reproduce macroscopic behavior. Classical data-driven techniques based on force-matching (multiscale coarse-graining) \cite{izvekov_forcematching2005} and relative entropy minimization \cite{shell_relative_cg_pot_2008} learn variational models that approximate the PMF. With the rise in deep learning, these methods have been combined with highly expressive neural networks, creating highly expressive CG potentials \cite{chen_deepCG_2018, wang_machine_2019, husic_coarse_2020, thaler_deep_2022}. In \citeauthor{kohler_flow-matching_2023}, the advantages of both force-matching and relative entropy are combined into a new training method called flow-matching \cite{kohler_flow-matching_2023}. Data-driven, generative models based on Variational Autoencoders (VAEs) \cite{kingma_auto-encoding_2013} or Generative Adversarial Networks (GANs) \cite{goodfellow2014generativeadversarialnetworks} are capable of learning CG representations  and a back-mapping  simultaneously \cite{wang_coarse-graining_2019, wang_generative_2022, li_backmapping_gan_2020, stieffenhofer2020adversarialreversemappingequilibrated, stieffenhofer2021gan}. 

The overwhelming majority of data-based techniques rely on first generating reference data based on long MD simulations, which are assumed to have captured all relevant modes in the configuration space. These are subsequently used to train the postulated CG model.
This creates a “chicken-and-egg” problem \cite{rohrdanz_discovering_2013}: CG models rely on all-atom simulation data to learn the system’s behavior, but apart from the insights they provide, they can at best reproduce what is already contained in the data. In other words, we need CG models to efficiently explore new configurations, yet we also need prior all-atom data to train them, creating a circular dependency that limits their ability to discover entirely new modes in the configuration space.
 Furthermore, we note that these two steps, i.e., data generation and learning, are generally detached from one another.
 Similar issues are encountered in the automatic discovery of collective variables (CVs) from  all-atom, simulation data. These are required by enhanced sampling techniques, such as umbrella sampling \cite{torrie_nonphysical_1977, souaille_extension_2001}, metadynamics \cite{barducci_metadynamics_2011, bonomi_reconstructing_2009}, or adaptive biasing potential methods \cite{laio_escaping_2002, bilionis_free_2012, ferguson_nonlinear_2011}, to bias all-atom simulations away from free-energy wells and explore the whole configurational space. Nevertheless, it is questionable whether the CVs discovered can lead to the discovery of other wells beyond those contained in the all-atom simulation data they were trained on.

An alternative to data-driven, bottom-up CG techniques is provided by energy-based methods. These aim to approximate the Boltzmann distribution $p(\bx)$ using only evaluations of the energy (or interatomic potential) $U(\bx) = -\beta^{-1} \log p(\bx)$ and its derivatives (i.e., interatomic forces). A prominent example is the family of deep generative models known as Boltzmann Generators (BGs) \cite{noe_boltzmann_2019}, which train normalizing flows (NFs) \cite{norm_flows2019} using a combination of data and energy. BGs have been shown to generalize across different thermodynamic states, such as temperatures and pressures \cite{tempBG2022, scalable_Schebek_2024}. Recent work on equivariant flows \cite{eqflows_kohler20a} has further demonstrated that incorporating symmetry-preserving architectures into BGs can significantly improve generalization and sampling efficiency. BGs are capable of generating one-shot, independent samples and obtaining unbiased estimates of observables through importance sampling. However, they operate entirely in the all-atom coordinate space and do not employ a coarse-grained description, which limits their scalability and interpretability.

Another line of research avoids reweighting and instead uses normalizing flows as proposals within subsequent Markov chain Monte Carlo (MCMC) schemes \cite{wu2020stochasticnormalizingflows, gabrie_adaptive_2022, samsonov2022, abflowMC2024}. Among these, the adaptive Monte Carlo framework of \citeauthor{gabrie_adaptive_2022} trains the NF on-the-fly during sampling, combining local MCMC updates with NF-based nonlocal moves. This hybrid strategy improves mixing efficiency and sampling performance but still relies on having initial data covering all relevant modes; otherwise, the NF may fail to learn unexplored regions of the distribution.
In contrast to these approaches, our framework introduces a fully generative, data-free method that operates in a structured latent space, enabling principled coarse-graining, automatic reconstruction of atomistic configurations, and accurate sampling of the Boltzmann distribution---all without requiring precollected trajectories.

 The simplest approach for pure energy-based training is minimizing the reverse Kullback-Leibler (KL) divergence. In \cite{wirnsberger_normalizing_2022} a normalizing flow model is trained to match the Boltzmann distribution of atomic solids with up to 512 atoms. However, without any dimensionality reduction, the use of this approach is computationally expensive. Also, it is known that the reverse KL-divergence suffers from a mode-seeking behavior \cite{stimper2022resamplingbasedistributionsnormalizing,felardos_designing_2023} which is a problem amplified in higher dimensions, even as  those encountered in  simple protein systems. In \cite{felardos2023designinglossesdatafreetraining}, the authors analyze mode collapse during optimization for small atomistic systems and suggest alternative training loss terms. However, they are only able to improve upon a pretrained model. Alternatively, researchers have tried using the $\alpha$-divergence, which exhibits better mass-covering properties \cite{midgley_flow_2023}. The authors additionally used Annealed Importance Sampling (AIS) to facilitate the discovery of new modes. They are the first to learn the Boltzmann distribution of a small protein, alanine dipeptide, purely from its unnormalized density. On the downside, AIS still requires significant computational resources, as mollified versions of the target Boltzmann distribution need to be sampled. 
 Most of these methods operate on global internal coordinates to reduce the complexity of the learning objective. In 
 \cite{midgley2024se3equivariantaugmentedcoupling}, equivariance is directly incorporated into the flow architecture through internal auxiliary variables, while still operating on Cartesian coordinates. 
 However, the equivariant layers are still expensive, and their models have not been applied to protein systems for pure energy training. Purely machine-learning-based neural samplers such as the Path Integral Sampler (PIS) \cite{zhang_path_2022}, Denoising Diffusion Sampler (DDS) \cite{vargas_denoising_2023}, Time-reversed Diffusion Sampler (DIS) \cite{berner_optimal_2024}, and Iterated Denoising Energy Matching (iDEM) \cite{akhound-sadegh_iterated_2024}, present powerful tools for approximating Boltzmann distributions without molecular dynamics data. While these methods amortize MCMC sampling and can be trained without trajectories, they operate in the full atomistic dimension and do not incorporate physical priors or coarse-graining structure. Moreover, methods relying on stochastic differential equations (SDEs) and diffusion processes often require architectural tricks (e.g., Langevin preconditioning) that hinder simulation-free learning and raise compatibility issues \cite{he_no_2025}.

The main  idea of this work is to reparameterize the full atomistic configuration $\bx$ using a bijective, learnable transformation that decomposes the system into two components: a set of coarse-grained variables $\bz$, and a complementary set of variables $\bX$. This decomposition is driven by statistical principles: the marginal distribution of $\bz$ is encouraged to be \textbf{multimodal}, capturing the metastable states one would encounter in molecular dynamics, while the conditional distribution of $\bX$ given $\bz$ is constrained to be \textbf{unimodal}, representing localized thermal fluctuations.
Among the infinitely many possible transformations, we seek one that naturally induces this statistical structure through the form of an approximating distribution. We model the joint density over $\bX$ and $\bz$ as a product of two terms: (i) a flexible, potentially multimodal marginal over $\bz$, parameterized via a normalizing flow, and (ii) a unimodal conditional over $\bX$ given $\bz$, such as a Gaussian. These properties are not enforced on the transformation itself but emerge naturally through the design of the learning objective.

To this end, we minimize the Kullback-Leibler divergence between the approximating distribution and the transformed Boltzmann distribution. This leads to the simultaneous achievement of two core objectives:
\begin{enumerate}
    \item Learn a coarse-graining transformation that captures the statistical structure of the system;
    \item Fit an expressive, generative probabilistic model that embeds coarse-graining behavior into its very architecture.
\end{enumerate}
Complementing the statistical formulation is a dynamical interpretation that provides further intuition. The coarse variables $\bz$ can be seen as capturing the system's ``slow'' degrees of freedom, while the ``fast'' variables $\bX$, conditioned on $\bz$, rapidly equilibrate. Although the method does not rely on dynamical data, this Perspective highlights the alignment between the statistical structure and physical behavior. 

This method provides several appealing features:
\begin{itemize}
    \item By approximating the full Boltzmann distribution in transformed coordinates, the model achieves thermodynamic consistency at both the coarse-grained and fine-grained levels \cite{noid_systematic_2013}. This implies that, up to approximation errors, the model can reproduce expectations of arbitrary observables \cite{chennakesavalu_ensuring_2023}.
    
    \item The bijective transformation offers a natural avenue to inject physical insight  into the selection of coarse-grained variables, thereby enhancing interpretability and generalization in alignment with physical intuition \cite{rotskoff_sampling_2024}.
    
    \item Unlike traditional coarse-graining techniques that struggle with the ill-posed inverse problem of back-mapping atomistic details onto coarse-grained configurations, our generative model directly addresses this challenge. Since the mapping is bijective and learned, one can reconstruct full-resolution atomistic configurations, overcoming the ``one-to-many'' ambiguity inherent in back-mapping \cite{christofi_physics-informed_2024}.

    \item Crucially, our framework does not require precollected MD trajectories to fit a coarse-grained model. Instead, training relies solely on evaluations of the all-atom force field, eliminating the need for costly and potentially biased data generation that hampers traditional approaches. 
    
\end{itemize}

Overall, this work introduces a principled, data-free, and physically grounded approach to coarse-grained molecular modeling. It leverages recent advances in probabilistic modeling and generative learning to construct scalable, interpretable, and thermodynamically faithful models of complex molecular systems. 

The rest of the paper is structured as follows: 
In Section \ref{sec:Method}, we provide a detailed description of the energy-based coarse-graining methodology, highlighting its key principles and algorithmic framework, while discussing comparisons with popular alternatives. We then demonstrate the effectiveness of this approach in Section \ref{sec:Example}, where we apply it to several model systems: an asymmetric double-well (DW) potential, a Gaussian mixture model (GMM), and the protein system of  alanine dipeptide. Finally, in Section \ref{sec:Conc}, we summarize the main findings of our study and discuss potential avenues for further improvements and enhancements of the method.

\section{Methodology}
\label{sec:Method}
Whether unraveling complex protein folding or magnetic spin interactions, these challenges hinge on the Boltzmann distribution, which serves as the fundamental bridge between interatomic potentials and the probability of microscopic configurations in equilibrium statistical mechanics.
The challenge in exploring Boltzmann densities arises from the presence of multiple modes whose locations are generally unknown a priori. As a result, standard MD tools become trapped for a large number of steps combined with the high dimensionality, this renders such calculations impractical or even impossible.

In the following, we propose a generative coarse-graining scheme requiring only evaluations of the interatomic potential and its gradient for training. The core assumption underlying all coarse-graining approaches is that the multimodality of the target Boltzmann distribution is concentrated in a significantly lower-dimensional subspace, or better yet manifold.
In a dynamical context, the coordinates along this manifold are often referred to as ``slow'' degrees of freedom, while the remaining ones are constrained by them \cite{hummer_coarse_2003}. In the statistical setting advocated in this work, the {\em marginal density} of the ``slow'' DOFs would still be multimodal (albeit living in lower dimensions), whereas the {\em conditional density} of the remaining DOFs would be unimodal (and possibly quite narrow).
This, in fact, serves as the overarching principle in the ensuing formulations.

\subsection{Probabilistic Generative Model}
\label{sec:ProbGen}

We now formalize the generative coarse-graining framework and describe how the target Boltzmann density is represented under a bijective transformation.
We consider an ensemble of $n$ atoms,  each of which has coordinates  $\bx^{(i)} \in \mathbb{R}^3, i=1,\ldots, n$ that are collectively represented with the vector $\bx \in \mathcal{M} \subset \mathbb{R}^{d_{\bx}}$ where $d_{\bx}=3n$. If $U(\bx)$ is the interatomic potential, then the target Boltzmann density is defined as:
 \begin{equation}
    p(\bx~;\beta) = \cfrac{ e^{-\beta U(\bx)} }{Z_{\beta}}, \label{eq:Bo}
 \end{equation}
where $\beta = 1/(k_B T)$ is the inverse temperature, $k_B$ the Boltzmann constant, $T$ the temperature, and $Z_{\beta}$ the partition function.

\begin{figure}[!t]
\includegraphics[height=0.3\textheight]{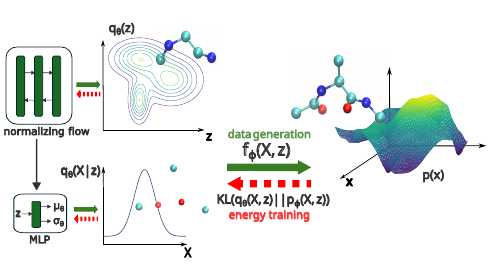}
\caption{Schematic illustration of the proposed {\em generative} framework. Two sets of latent coordinates are identified: a)  $\bz$ with a {\em multimodal}, learnable density $q_{\bt}(\bz)$ corresponding to ``slow'' DOFs, and b)  $\bX$  with a {\em unimodal}, learnable, conditional  density $q_{\bt}(\bX|\bz)$ corresponding to ``fast'' DOFs modulated by $\bz$. These are combined in order to reconstruct the all-atom DOFs $\bx$ through the learnable diffeomorphism $\bs{f}_{\bphi}$.}
\label{fig:overview}
\end{figure}

We introduce two new sets of DOFs, the meaning of which is explained in the sequel, namely $\bX \in \mathcal{X}$ and $\bz \in  \mathcal{Z}$ through a potentially nonlinear, differentiable,  bijective, parameterized mapping:
\be
\bx=\bs{f}_{\bphi}(\bX,\bz)
\label{eq:diffeo}
\ee
where $\bs{f}_{\bphi}: \mathcal{X} \times \mathcal{Z} \to \mathcal{M}$ and $\bphi$ are the associated parameters. We provide specific forms for $\bs{f}_{\bphi}$ in the sequel.
As illustrated in Figure~\ref{fig:overview}, the ``slow'' variables $\bz$ are modeled with a flexible, potentially multimodal density $q_{\bt}(\bz)$, while the ``fast'' variables $\bX$ are modeled conditionally, with a simple unimodal density $q_{\bt}(\bX|\bz)$. Together, they are mapped  to the all-atom coordinates $\bx$ through the learnable bijective transformation $\bs{f}_{\bphi}$.

Since $\dim(\bx)=\dim(\bz)+\dim(\bX)$ the partition of the arguments of $\bs{f}_{\bphi}$ requires only deciding a priori about $\dim(\bz)$. 
We note that for a system for which the user has no prior knowledge, $\dim(\bz)$ will need to be manually adjusted based on model results. An automatic way to score and potentially refine trained models based on $\dim(\bz)$ is discussed in the Conclusions (Section \ref{sec:Conc}).
The corresponding density in the $\mathcal{X} \times \mathcal{Z}$-space would be:
\be
p_{\bphi}(\bX,\bz~;\beta)=\frac{ e^{-\beta U(\bs{f}_{\bphi}(\bX,\bz))}}{Z_{\beta}} K_{\bphi}(\bX,\bz)
\ee
where $K_{\bphi}(\bX,\bz)=\left| \det\left( \frac{\partial \bs{f}_{\bphi}}{\partial (\bX,\bz)} \right) \right|$.
 This can also be written as:
\be
p_{\bphi}(\bX,\bz~;\beta) = \frac{1}{Z_{\beta}} e^{-\beta U_{\bphi}(\bX,\bz~;\beta)}
\label{eq:pphi}
\ee
where:
\be
U_{\bphi}(\bX,\bz~;\beta) =U(\bs{f}_{\bphi}(\bX,\bz))-\beta^{-1} \log K_{\bphi}(\bX,\bz),
\label{eq:uphi}
\ee
i.e., the target density is $\bphi$-dependent.

An infinite number of such transformations arise by varying $\dim(\bz)$ and the parameters $\bphi$. We posit that a good set of coarse variables $\bz$ should ensure that the \textbf{conditional density} $p_{\bphi}(\bX \mid \bz)$ is \textbf{unimodal}, which naturally induces a \textbf{multimodal marginal density} $p_{\bphi}(\bz) = \int p_{\bphi}(\bX,\bz~;\beta) \, d\bX$ that captures the metastable states of the system. 
This separation between multimodal and unimodal components underpins the entire generative formulation.
From a dynamical point of view, when simulating $(\bX,\bz)$-coordinates in an MD/MCMC setting, one would observe that $\bX$ would be the ``fast'' DOFs, which are quickly constrained by $\bz$, whereas the latter would be the ``slow'' variables exhibiting  similar metastable features  as $\bx$, albeit in a  space of reduced dimension (Figure \ref{fig:overview}).

In order to discover a transformation that ensures the aforementioned properties, we consider an approximation to $p_{\bphi}(\bX,\bz~;\beta)$ of the form:
\be
q_{\bt}(\bX,\bz) = q_{\bt}(\bz)~q_{\bt}(\bX|\bz)
\ee
Based on the aforementioned objectives, we postulate that:
\begin{itemize}
\item $q_{\bt}(\bX|\bz)$ is a \textbf{unimodal} density (e.g., a Gaussian) and 
\item $q_{\bt}(\bz)$ is a potentially \textbf{multimodal} density. We model this implicitly by employing a normalizing flow  \cite{norm_flows2019} induced by the transformation $g_{\bt}(\bepsilon)$ such that: 
\be
\bz=g_{\bt}(\bepsilon), \label{eq:repz}
\ee
where $q(\bepsilon)$ is the standard Gaussian. As a result:
\be
\log q(\bepsilon)=\log q_{\bt}(g_{\bt}(\bepsilon))+\log J_{\bt}(\bepsilon)
\label{eq:qflow}
\ee
where $J_{\bt}(\bepsilon)=| \det( \frac{ \pa g_{\bt} }{ \pa \bepsilon})|$.
\end{itemize}

Having specified the model structure, we now turn to the learning objective. To learn both the transformation $\bs{f}_{\bphi}$ and the generative approximation $q_{\bt}$, we minimize the reverse Kullback–Leibler divergence, which encourages $q_{\bt}$ to match the transformed Boltzmann density $p_{\bphi}$.  Intuitively, this objective ensures that the learned model both identifies a physically meaningful decomposition into ``slow'' and ``fast'' variables and provides a powerful generative sampler. In particular, the objective $\mathcal{L}$ is:
\begin{align*}
\mathcal{L}(\bt,\bphi) & =KL(q_{\bt}(\bX,\bz)|| p_{\bphi}(\bX,\bz~;\beta))  \\
& = -\left<\log p_{\bphi}(\bX,\bz~;\beta)\right>_{q_{\bt}(\bX,\bz)}+\left<\log q_{\bt}(\bX,\bz) \right>_{q_{\bt}(\bX,\bz)} \\
& =  \beta \left< U_{\bphi}(\bX,\bz~;\beta) \right>_{q_{\bt}(\bX,\bz)}+\log Z_{\beta}+\left< \log q_{\bt} (\bX|\bz)\right>_{q_{\bt}(\bX,\bz)}+\left< \log ~q_{\bt}(\bz) \right>_{q_{\bt}(\bz)} \addtocounter{equation}{1}\tag{\theequation}
\label{eq:klflow}
\end{align*}

We note that in the general case, minimizing the aforementioned KL-divergence simultaneously achieves two objectives:
\begin{itemize}
\item learns a diffeomorphic transformation (through $\bphi$) which defines the coarse variables 
$\bz$ and the reconstruction map.
\item learns  the generative model 
 $q_{\bt}$,  which produces equilibrium-consistent samples in transformed coordinates.
\end{itemize}
We further note that the transformation $\bs{f}_{\bphi}$ does not involve dimensionality reduction and once learned can be readily used  in order to reconstruct the full atomistic picture, i.e., $\bx$. In combination with $q_{\bt}$ and to the extent this provides a good approximation to the transformed Boltzmann, one can therefore obtain one-shot samples  (see Section \ref{sec:predictions}).

Importantly, training requires only evaluations of the interatomic potential $U(\bx)$  (or equivalently $U_{\bphi}$) and its gradient (i.e., interatomic forces) \cite{noe_boltzmann_2019, schoebel_embedded_2020}. This energy-based training \cite{akhound-sadegh_iterated_2024}  does not require  MD trajectories or pregenerated data sets, directly overcoming the traditional ``chicken-and-egg problem'' faced by data-driven coarse-graining methods.

We consider a particular form of  such a bijective map $\bs{f}_{\bphi}$ which is linear, i.e.,
\begin{equation*}
\bx=\bs{A}_{\bphi} \left[ \begin{array}{c} \bz \\ \bX \end{array}\right]
\end{equation*}
 ($d_{\bx}=3n=\dim(\bx)$, $d_{\bz}=\dim(\bz)$ and $\dim(\bX)=d_{\bx}-d_{\bz}$) where for each atom $i$ with coordinates $\bx_{(i)}$ we have:
\be
\bx_{(i)}=\sum_{j=1}^{d_{\bz}} a_{\bphi_{i,j}} \bs{I} \bz_{(j)}+\sum_{j=d_{\bz}+1}^{d_{\bx}} a_{\bphi_{i,j}} \bs{I} \bX_{(j-d_{\bz})}
\ee
Here, $\bs{I}$ denotes the $3\times 3$ identity matrix, and $\bz_{(j)}$ and $\bX_{(k)}$ are the coordinates of pseudo-$\bz$-atom $j$ and pseudo-$\bX$-atom $k$, respectively. One can readily show that if:
\be
\sum_{j=1}^{d_{\bx}} a_{\bphi_{i,j}} =1, \forall i
\label{eq:Arow}
\ee
then the corresponding map is \textbf{equivariant} to rigid-body motions \cite{wang_coarse-graining_2019}. The associated $\bs{A}_{\bphi}$ matrix is a right stochastic matrix. We note that typical CG techniques, which lump atoms into bigger, pseudo/virtual-atoms, arise by particular choices of the coefficients $a_{\bphi_{i,j}}$ \cite{kalligiannaki_geometry_2015,sahrmann_utilizing_2023}. 
Unlike these methods, however, our approach learns the mapping and, crucially, retains and models the additional DOFs (i.e., $\bX$), enabling a full reconstruction (back-mapping) of the all-atom coordinates $\bx$. We note that in this case  $K_{\bphi}(\bX,\bz)=\left| \det\left( \frac{\partial \bs{f}_{\bphi}}{\partial (\bX,\bz)} \right) \right|=|\det(\bs{A}_{\bphi})|$, which is independent of $\bX,\bz$.

A special case of the aforementioned linear map is when $\bs{A}_{\bphi}$ is a {\em  permutation matrix} which arises when $a_{\bphi_{i,j}}=0 \textrm{ or }1$ and there is a single $1$ per row and column (in this case $K_{\bphi}(\bX,\bz)=1$). Such a map implies a partition of all-atom coordinates $\bx$. An illustration of such a partitioning can be seen in Figure \ref{fig:overview} for the alanine dipeptide, where $\bz$ represents the coordinates of actual, backbone atoms and $\bX$ represents the coordinates of the side-chain atoms.

\noindent \textbf{Remarks:}
\begin{itemize}
\item Alternative learning objectives, such as the Fisher divergence \cite{hyvarinen_estimation_2005} or the $\chi^2$-divergence \cite{dieng_variational_2017,midgley_flow_2023}, which have been employed in the past and have shown advantages over the reverse KL-divergence adopted herein, could be readily used, but are not pursued in this study.  In the subsequent section, we discuss in detail how this KL-divergence can be minimized and the parametrization adopted for the approximation $q_{\bt}$.

\item It is instructive to  compare the proposed  method with the popular, relative entropy (RE) method~\cite{shell_relative_cg_pot_2008}, which employs the \emph{forward} KL-divergence in the context of coarse-graining. Assuming a coarse-grained map $\mathcal{P}$, which can be thought of as a partial inverse of $\bs{f}_{\bphi}$ in~\rfeq{eq:diffeo}, onto the same, lower-dimensional space $\bz = \mathcal{P}(\bx)$ and the same CG model $q_{\bt}(\bz)$, the RE method minimizes the KL-divergence between the (intractable) marginal Boltzmann density of the CG coordinates,
\[
p(\bz) = \int \delta(\bz - \mathcal{P}(\bx)) ~ p(\bx) ~ d\bx,
\]
and its approximant $q_{\bt}(\bz)$, as
\[
S_{\mathrm{rel}}(\bt) = \mathrm{KL}(p(\bz) \,\|\, q_{\bt}(\bz)) = \left\langle \log p(\bz) - \log q_{\bt}(\bz) \right\rangle_{p(\bz)},
\]
where $S_{\mathrm{rel}}(\bt) \geq 0$ by Gibbs' inequality, with equality if and only if $p(\bz) = q_{\bt}(\bz)$ almost everywhere.

During optimization, the first term $\langle \log p(\bz) \rangle_{p(\bz)}$ can be neglected, as it is independent of the model parameters $\bt$. Exploiting the definition of $p(\bz)$, the objective can be rewritten as an expectation over $p(\bx)$:
\[
S_{\mathrm{rel}}(\bt) = \left\langle - \log q_{\bt}(\mathcal{P}(\bx)) \right\rangle_{p(\bx)}.
\]
Thus, minimizing the relative entropy corresponds to maximizing the likelihood that the CG model $q_{\bt}(\bz)$ reproduces the statistics of the mapped atomistic system, emphasizing coverage of all relevant modes (i.e., mass-covering behavior) rather than focusing on the dominant ones.

However, evaluating this expectation requires sampling from $p(\bx)$, which is generally intractable and must be approximated via all-atom MD or MCMC simulations. Consequently, the performance of RE-based coarse-graining is fundamentally limited by the quality and completeness of the available data. Furthermore, even if $q_{\bt}(\bz)$ provides an excellent approximation to $p(\bz)$, it does not inherently solve the reconstruction (or back-mapping) problem: additional, often heuristic, steps are necessary to generate consistent all-atom configurations $\bx$ from a given $\bz$.
\end{itemize}

\subsection{Training Framework}
\label{sec:training}
In this section, we describe the algorithmic steps for training the proposed model, along with a tempering scheme designed to address known challenges in minimizing the reverse KL-divergence \cite{felardos_designing_2023}. We also provide details on parameters $\bt$ and explain why lightweight normalizing flow models can be effective in our formulation.

Given the unimodality assumption for  $q_{\bt}(\bX|\bz)$, we employ a  Gaussian:
\be
q_{\bt}(\bX|\bz) = \mathcal{N}(\bX|\mu_{\bt}(\bz), \mathrm{diag}(\sigma^2_{\bt}(\bz)))
\label{eq:cond}
\ee
with mean $\mu_{\bt}(\bz)$ and a diagonal covariance matrix with variances given by the vector $\sigma_{\bt}^2(\bz)$. This is convenient as it leads to direct reparametrization in the form: 

\be
\bX = \mu_{\bt}(\bz) + \sigma_{\bt}(\bz) \bepsilon_{\bX} = h_{\bt}(\bepsilon_{\bX}, \bz),
\label{eq:repX}
\ee
where $\bepsilon_{\bX} \sim q(\bepsilon_{\bX})=\mathcal{N}(\bs{0}, \bs{I})$.

Following \rfeq{eq:klflow} and using the reparametrizations implied by the normalizing flow of \rfeq{eq:repz} and in \rfeq{eq:repX}, the learning objective $\mathcal{L}(\bt,\bphi)$ can be written as:

\begin{align*}
\mathcal{L}(\bt,\bphi) & = \beta \left< U_{\bphi}(\bX,\bz~;\beta) \right>_{q_{\bt}(\bX,\bz)} + \left< \log q_{\bt} (\bX|\bz)\right>_{q_{\bt}(\bX,\bz)}+\left< \log ~q_{\bt}(\bz) \right>_{q_{\bt}(\bz)} \\
& = \beta \left< U_{\bphi}(h_{\bt}(\bepsilon_{\bX}, g_{\bt}(\bepsilon)),g_{\bt}(\bepsilon)) \right>_{q(\bepsilon,\bepsilon_{\bX})} + \left< \left < \log q_{\bt} (\bX|g_{\bt} (\bepsilon)) \right>_{q_{\bt}(\bX |g_{\bt}(\bepsilon))} \right>_{q(\bepsilon)}\\
& +\left<\log q_{\bt}(g_{\bt}(\bepsilon)) \right>_{q(\bepsilon)},\numberthis
\label{eq:klmarg}
\end{align*}

where the partition function is independent of $(\bt, \bphi)$ and can therefore be ignored during optimization.

In view of \rfeq{eq:cond}, the second term can be expressed (ignoring constants) as
\be
\left< \left < \log q_{\bt} (X|g_{\bt} (\bepsilon)) \right>_{q_{\bt}(\bX|g_{\bt}(\bepsilon))} \right>_{q(\bepsilon)} = -\frac{1}{2} \left< \log ~\det(\mathrm{diag}(\sigma^2_{\bt}(g_{\bt}(\bepsilon)))) \right>_{q(\bepsilon)}
\ee

We note that the third term (with the help of \rfeq{eq:qflow}) can be rewritten as:
\be
\left<\log q_{\bt}(g_{\bt}(\bepsilon)) \right>_{q(\bepsilon)} = \left< \log q(\bepsilon) \right>_{q(\bepsilon)} - \left< \log J_{\bt}(\bepsilon) \right>_{q(\bepsilon)}, \label{eq:flow}
\ee
derivatives of which with respect to $\bt$ can be readily obtained from the Jacobian $J_{\bt}$ of the flow.

Gradients of the objective, which are used for its minimization, can be obtained through the application of the chain rule. In particular and with respect to $\bt$, we have:

\begin{align*}
\nabla_{\bt} \mathcal{L}(\bt,\bphi)   &= \beta \left< \frac{\pa U_{\bphi}}{\pa \bX} \left( \frac{\pa h_{\bt}}{\pa \bt} + \frac{\pa h_{\bt}}{\pa \bz}\frac{\pa g_{\bt}}{\pa \bt} \right) + \frac{\pa U_{\bphi}}{\pa \bz} ~\frac{\pa g_{\bt}}{\pa \bt} \right>_{q(\bepsilon,\bepsilon_{\bX})}
 - \frac{1}{2} \left< \frac{\pa \log ~\det( \mathrm{diag}(\sigma_{\bt}(g_{\bt}(\bepsilon))))}{\pa \bt} ~\right>_{q(\bepsilon)}\\
 & - \left< \frac{\pa \log J_{\bt}(\bepsilon)}{\pa \bt} ~\right>_{q(\bepsilon)} \numberthis
 \label{eq:gradtheta}
\end{align*}

The gradient with respect to $\bphi$ is only dependent on the first term of \rfeq{eq:klmarg} and in view of \rfeq{eq:uphi} can be written as:
\begin{align*}
\nabla_{\bphi} \mathcal{L}(\bt,\bphi)  &= \frac{\pa}{\pa \bphi}\beta \left< U_{\bphi}(h_{\bt}(\bepsilon_{\bX}, g_{\bt}(\bepsilon)),g_{\bt}(\bepsilon)) \right>_{q(\bepsilon,\bepsilon_{\bX})} \\
&= \beta \left< \frac{\pa U}{\pa \bx}\frac{\pa \bs{f}_{\bphi} (h_{\bt}(\bepsilon_{\bX}, g_{\bt}(\bepsilon)),g_{\bt}(\bepsilon)) }{\pa \bphi}  - \frac{1}{\beta}\frac{\pa \log K_{\bphi}(h_{\bt}(\bepsilon_{\bX}, g_{\bt}(\bepsilon)),g_{\bt}(\bepsilon)) }{\pa \bphi} \right>_{q(\bepsilon,\bepsilon_{\bX})} \numberthis
\label{eq:gradphi}
\end{align*}

For a given value of the model parameters $(\bt,\bphi)$, Monte Carlo estimates of the aforementioned gradients, which can be used during training, can be obtained by following the steps below:

\begin{enumerate}
\item Generate $N$ independent samples $\{\bepsilon^{(i)}\}_{i=1}^N$, $\{\bepsilon_{\bX}^{(i)}\}_{i=1}^N$  from the base, standard Gaussian  densities $q(\bepsilon), q(\bepsilon_{\bX})$.
\item For each such sample-pair $i$, compute the:
\begin{itemize}
\item $\bz$-coordinates as $\bz^{(i)}=g_{\bt}(\bepsilon^{(i)}), ~i=1,\ldots,N$, 
\item $\bX$-coordinates as $\bX^{(i)}=h_{\bt}(\bepsilon_{\bX}^{(i)},\bz^{(i)}), ~i=1,\ldots,N$, and
\item $\bx$-coordinates as $\bx^{(i)}=\bs{f}_{\bphi}(\bX^{(i)},\bz^{(i)})$.
\end{itemize}
\item Compute  forces on the $\bx$-atoms $\bs{F}_{\bx}^{(i)}=-\nabla_{\bx}U(\bx^{(i)})$.
\item Compute the forces on the $\bz$-atoms and $\bX$-atoms :
\begin{align*}
\bs{F}_{\bz}^{(i)}=\bs{F}_{\bx}^{(i)}~\nabla_{\bz}\bs{f}_{\bphi}(\bX^{(i)},\bz^{(i)}) + \beta^{-1}~\nabla_{\bz} \log K_{\bphi}(\bX^{(i)},\bz^{(i)})\\
\bs{F}_{\bX}^{(i)}=\beta\bs{F}_{\bx}^{(i)}~\nabla_{\bX}\bs{f}_{\bphi}(\bX^{(i)},\bz^{(i)}) + \beta^{-1}~\nabla_{\bX} \log K_{\bphi}(\bX^{(i)},\bz^{(i)}) \label{eq:forcesXz} \numberthis
\end{align*}
\item Approximate $\bt$-gradient (see \rfeq{eq:gradtheta}):
\begin{align*}
\nabla_{\bt} \mathcal{L}(\bt,\bphi)    &\approx \frac{1}{N} \sum_{i=1}^N ~  -\beta \bs{F}_{\bX}^{(i)}  \left( \nabla_{\bt}h_{\bt}(\bepsilon_{\bX}^{(i)},\bz^{(i)}) + \nabla_{\bz}h_{\bt}(\bepsilon_{\bX}^{(i)},\bz^{(i)})~\nabla_{\bt}g_{\bt}(\bepsilon^{(i)}) \right) \\
&- \beta \left( \bs{F}_{\bz}^{(i)} ~\nabla_{\bt}g_{\bt}(\bepsilon^{(i)}) \right) -\frac{1}{2} \nabla_{\bt}\log ~\det(\mathrm{diag}(\sigma^2_{\bt}(\bz^{(i)}))) - \nabla_{\bt}\log J_{\bt}(\bepsilon^{(i)}) \numberthis
\label{eq:gradthetamc}
\end{align*}
\item Approximate $\bphi$-gradient (see \rfeq{eq:gradphi}):
\be
\nabla_{\bphi} \mathcal{L}(\bt,\bphi)   \approx \frac{1}{N} \sum_{i=1}^N ~   -\beta \bs{F}_{\bx}^{(i)} ~\nabla_{\bphi}\bs{f}_{\bphi}(\bX^{(i)},\bz^{(i)}) - \nabla_{\bphi}\log K_{\bphi}(\bX^{(i)},\bz^{(i)})
\label{eq:gradphimc}
\ee
\end{enumerate}

\noindent \textbf{Remarks:}
\begin{itemize} 

\item The calculation above of the forces $\bs{F}_{\bz}^{(i)}$ and $\bs{F}_{\bX}^{(i)}$ ensures that we do not evaluate the potential energy multiple times, which is often the most  costly aspect of the gradient evaluation. We  further note that the second term in \rfeq{eq:forcesXz} vanishes for transformations of the form $\bx = \bs{A}_{\bphi} \left[ \begin{array}{cc} \bz & \bX \end{array} \right]^{\!\top}$, which we introduced earlier as the Jacobian $K_{\bphi}(\bX,\bz)=|\det(\bs{A}_{\bphi})|$ is independent of $\bX$ and $\bz$.

\item The proposed method shares some similarities with Boltzmann Generators, as it uses normalizing flows to approximate the Boltzmann distribution. However, there are some key differences. Boltzmann Generators\cite{noe_boltzmann_2019} combine data-based and energy-based training of normalizing flows with reweighting to obtain unbiased samples from the target Boltzmann distribution. In contrast, we discover a density $q_{\bt}(\bz)$ that lives in a lower-dimensional space compared to the original Boltzmann distribution $p(\bx)$ but nevertheless encompasses the multiple modes that are present in the latter and which are a priori unknown. We do so by stabilizing energy training with an adaptive tempering scheme. Some BGs employ a partitioning of coordinates (e.g., backbone and side chain atoms) in the neural network architecture of the flow model and in the construction of the internal coordinates in order to improve training efficiency. Our method builds the partition of ``slow'' and ``fast'' DOFs directly into the probabilistic framework in the form of a unimodal conditional distribution $q_{\bt}(\bX|\bz)$. This simplifies the learning process and pushes the multimodality into the aforementioned density $q_{\bt}(\bz)$.

\item The Monte Carlo estimates of the gradients of the training objective in Equations (\ref{eq:gradthetamc}) and (\ref{eq:gradphimc})
 are used to update the parameters ($\bt,\bphi$) using a Stochastic Gradient Descent (SGD) scheme. In particular, we use the ADAM optimizer \cite{Kingma2015} with parameters $\beta_1=0.99$, $\beta_2=0.999$, and $\epsilon_{\mathrm{ADAM}}=1.0 \times 10^{-8}$.
\end{itemize}

The minimization of the reverse KL-divergence as in \rfeq{eq:klmarg} is fraught with well-documented computational difficulties \cite{bishop2007, shu_amortized_2018, missmodes2020}. In particular, it exhibits a {\em mode-seeking} behavior, which in the context of multimodal target densities considered, can be particularly deleterious as it can lead to approximations $q_{\bt}(\bX,\bz)$ that miss some important mode(s) \cite{felardos_designing_2023}. The reverse KL-divergence penalizes $q_{\bt}$ for placing probability mass, where $p_{\bphi}$ is small. As a result, $q_{\bt}$ tends to concentrate on regions where $p_{\bphi}$ is large, while avoiding areas of low support. This exclusion of low-density regions of $p_{\bphi}$ gives rise to its mode-seeking behavior \cite{minka2005divergence}. The most important mitigating factor in the proposed formulation, as compared to others that have used the reverse KL \cite{wirnsberger_normalizing_2022, noe_boltzmann_2019, felardos2023designinglossesdatafreetraining}, is that training is carried out in a (much) lower-dimensional space $\mathcal{Z}$ as compared to the original Boltzmann.
The second mitigating factor is a tempering scheme that we employ, the effectiveness of which is illustrated empirically in the numerical experiments (Section \ref{sec:DoubleWell}). 
We note that for  $\beta \to 0$ (or equivalent $T \to \infty$), the target Boltzmann is effectively uniform and unimodal.
As $\beta$ is slowly increased, the modes become more pronounced, but as long as this is done carefully, the approximation can track them. While the shape of the modes may change, updates to $\bt$ can easily account for it. 
Furthermore, the approximation obtained at a certain $\beta$ serves as a good initial guess for subsequent $\beta$. 
An additional benefit of such a strategy is that one obtains a CG generative model for all intermediate $\beta$ values considered. 

To this end, we employ an adaptive, information-theoretic scheme that automatically determines the sequence of $\beta$ values, starting from $0$ and progressing to the target $\beta_{\mathrm{target}}$, ensuring a smooth transition that captures all relevant modes \cite{schoebel_embedded_2020}.
Let $q_{\bt}(\bX,\bz)$ be the optimal approximation to the target $p_{\bphi}(\bX,\bz~;\beta_k)$ for the current $\beta=\beta_k$ at step $k$. Our goal is to determine $\beta_{k+1}=\beta_k+\Delta \beta_k$,  i.e.,  to identify $\Delta \beta_k >0$
 so as $p_{\bphi}(\bX,\bz~;\beta_{k+1})$ does not differ substantially from $p_{\bphi}(\bX,\bz~;\beta_k)$ and one can readily transition to the new optimal approximation $q_{\bt}$. We note that even though the parameters $\bphi$ are also updated at the new $\beta_{k+1}$ in order to find their new optimal values, this is not considered in the adaptivity metrics.

For this purpose, we employ the relative change in the KL-divergence used as the learning objective, namely:
\be
\delta KL_k (\Delta \beta_k)= \frac{KL(q_{\bt}(\bX,\bz)|| p_{\bphi}(\bX,\bz;~\beta_{k+1})) - KL(q_{\bt}(\bX,\bz)|| p_{\bphi}(\bX,\bz;~\beta_k))}{KL(q_{\bt}(\bX,\bz)|| p_{\bphi}(\bX,\bz;~\beta_k))}
\label{eq:dkl}
\ee
We note that when $\beta_{k+1} \to \beta_k$, $\delta KL_k \to 0$, and as $\beta_{k+1}$ deviates from $\beta_k$, we would expect $\delta KL_k $ to increase.  Hence, we define an upper bound $\delta KL_{\max}$ and select $\Delta \beta_k$ such that
\be
\Delta \beta_k = \min \{ \delta KL_k (\Delta \beta_k) =\delta KL_{\max}, ~\Delta \beta_{\max}, \beta_{\mathrm{\mathrm{target}}}-\beta_k \}
\label{eq:deltabetak}
\ee
where $\Delta \beta_{\max}$ is another user-defined threshold and $\beta_{\mathrm{target}}$ is the maximum $\beta$ (or equivalently, minimal absolute temperature) of interest. Values for the parameters of \rfeq{eq:deltabetak} are reported in Section \ref{sec:Example}. In the Appendix \ref{appendix:adap}, we provide details regarding the numerical approximation of $\delta KL_k$, which is based on Importance Sampling.

We finally provide an overview of the data-free training of the generative model proposed in pseudo-Algorithm~\ref{alg:upd}, where the outer loop increments $\Delta \beta_k$ according to the adaptive scheme described above, while the inner loop updates the model parameters $\bt$ and $\bphi$ using Stochastic Gradient Descent and the aforementioned, Monte Carlo estimates of the gradients.
A combination of gradient norm reduction, loss change, and maximum iterations was used to detect convergence.

\begin{algorithm}[!t]
\begin{spacing}{1.4}
\caption{Training algorithm with adaptive tempering}
\begin{algorithmic}[1]
\State \textbf{Input:} Target inverse temperature $\beta_{\mathrm{target}}$, maximal relative KL increase $\delta KL_{\max}$, maximal $\beta$-increment $\Delta \beta_{\max}$, learning rate $\eta_{\mathrm{SGD}}$.

\State \textbf{Initialize:} Parameters $(\bt,\bphi)$, $\beta_0 \gets 0$, adaptive tempering counter $k \gets 0$.
\While{$\beta_k < \beta_{\mathrm{target}}$} \Comment{Outer tempering loop}
\While{not converged} \Comment{Inner optimization loop}

    \State Compute  Monte Carlo estimates of gradients $\nabla_{\bt} \mathcal{L}$ and $\nabla_{\bphi} \mathcal{L}$ \par \Comment{see Equations~(\ref{eq:gradthetamc}) and (\ref{eq:gradphimc})}  
    \State Update parameters: $\bt \gets \bt + \eta_{\mathrm{SGD}} \cdot \nabla_{\bt} \mathcal{L}$
    \State Update parameters: $\bphi \gets \bphi + \eta_{\mathrm{SGD}} \cdot \nabla_{\bphi} \mathcal{L}$
\EndWhile
\State Compute increment $\Delta \beta_k$ \Comment{see \rfeq{eq:deltabetak}}
\State Update inverse temperature: $\beta_{k+1} \gets \beta_k + \Delta \beta_k$
\State Increment counter: $k \gets k + 1$
\EndWhile
\end{algorithmic}
\label{alg:upd}
\end{spacing}
\end{algorithm}

\subsection{Predictions}
\label{sec:predictions}
Once our model is fully trained, we can generate one-shot samples, which approximately follow the target Boltzmann distribution with the following steps:

\begin{enumerate}
\item Generate $N$ samples $\{\bepsilon^{(i)}\}_{i=1}^N$ from the base density $q(\bepsilon)$.
\item Compute the $\bz$-coordinates as $\bz^{(i)}=g_{\bt}(\bepsilon^{(i)}), ~i=1,\ldots,N$.
\item Sample the conditional $\bX^{(i)} \sim q_{\bt}(\bX|\bz^{(i)})$.
\item Transform back to the $\bx$-coordinates as $\bx^{(i)}=\bs{f}_{\bphi}(\bX^{(i)},\bz^{(i)})$.
\end{enumerate}

Furthermore, we can evaluate the free energy $A(\bz)=-\beta^{-1} \log q_{\bt}(\bz)$ to calculate transition paths and energy differences in the latent $\mathcal{Z}$-space. Moreover, we do not only obtain one model for the target Boltzmann distribution, but for each intermediate distribution chosen during tempering. This allows us to generate samples at each $\beta_k$ for which we converged during training.
We  note that $q_{\bt}(\bz)$ provides, in essence, a thermodynamically consistent projection of the original Boltzmann distribution, which can be further processed in order to learn, e.g., collective variables \cite{rohrdanz_discovering_2013,siddiqui_application_2023} or as the starting point for further coarse-graining operations, which can proceed in a hierarchical fashion.
Unbiased estimates of any physical observable $a(\bx)$ with respect to the Boltzmann density $p(\bx; \beta)$, and for any $\beta$ in the sequence considered during training, can be calculated using Importance Sampling as:
\begin{align*}
\left< a(\bx) \right>_{p(x; \beta)}  & = \int a(\bx) ~p(\bx; \beta) ~d\bx \\
 & = \int a(\bx) ~\frac{e^{-\beta U(\bx; \beta)}}{Z_{\beta}} ~d\bx \\
 & = \int a(\bs{f}_{\bphi}(\bX,\bz))~\frac{e^{-\beta U_{\bphi}(\bX,\bz~;\beta)}}{Z_{\beta}} ~d\bX~d\bz \\
 & = \int a(\bs{f}_{\bphi}(\bX,\bz))~w(\bX,\bz) ~q_{\bt}(\bX,\bz) ~d\bX~d\bz \\
 & \approx \sum_{m=1}^M W^{(m)} a(\bs{f}_{\bphi}(\bX^{(m)},\bz^{(m)})), \qquad (\bX^{(m)},\bz^{(m)})\sim q_{\bt}(\bX,\bz), \numberthis
\end{align*}
where $w(\bX,\bz)= \frac{1}{Z_{\beta}} \frac{ e^{-\beta U_{\bphi}(\bX,\bz; \beta)} }{q_{\bt}(\bX,\bz)}$ and $W^{(m)}$ are the normalized IS weights computed as $W^{(m)}=\frac{w^{(m)}}{\sum_{m'=1}^M w^{(m')} }$ using the unnormalized weights $w^{(m)}=\frac{ e^{-\beta U_{\bphi}(\bX^{(m)},\bz^{(m)}; ~\beta)} }{q_{\bt}(\bX^{(m)},\bz^{(m)})}$.
Furthermore, as long as the dominant modes have been captured by $q_{\bt}$, it could serve as the starting point in a bridging density, e.g.,  $q_{\bt}^{(1-\gamma)} ~p_{\bphi}^{\gamma}, ~\gamma \in [0,1]$  that would quickly be explored with the help of Annealed Importance Sampling (AIS) \cite{neal_annealed_2001}, or even better, Sequential Monte Carlo (SMC) \cite{del_moral_sequential_2006}.

\subsection{Model Specification}
\label{sec:Model}

The basis of the formulation is the density  $q_{\bt}(\bz)$ with respect to the CG DOFs $\bz$, which should exhibit the requisite expressivity in order to adapt to the target marginal $p_{\bphi}(\bz)$.
It is vital that the model is capable of capturing all the different metastable states of the system and, therefore, has to be able to account for the multimodality in the energy landscape.
A popular choice to model an arbitrary multimodal distribution is a normalizing flow \cite{norm_flows2019}. These combine a sequence of bijective, deterministic transformations to convert a simple base distribution into any complex distribution, as shown in \rfeq{eq:repz}.

We use coupling layers on the Cartesian coordinates of the system, similar to Real NVP \cite{realnvp16} but substitute the affine layers with monotonic rational-quadratic splines \cite{durkan2019neuralsplineflows}. These splines have fully differentiable and invertible mappings, while allowing highly expressive transformations. This makes them a perfect candidates to capture complex multimodal distributions. They have been used in many different forms in the context of Boltzmann generators and normalizing flows for Boltzmann distributions \cite{energymaps2020, stimper2022resamplingbasedistributionsnormalizing, midgley2024se3equivariantaugmentedcoupling, kim2024scalablenormalizingflowsenable}. We emphasize, however, that the strength of the proposed framework primarily draws from the bijective decomposition and the projection of the multimodality onto reduced coordinates $\bz$. As a result, we can employ a much more lightweight and smaller neural network architecture, reducing the computational effort during training. It would be interesting to apply our methods with SE(3) equivariant coupling flows \cite{midgley2024se3equivariantaugmentedcoupling} as incorporating symmetries into the model directly can improve training efficiency and generalization \cite{steerable_cnns_choen2016, batzner_e3-equivariant_2022, eqflows_kohler20a, klein2023equivariantflowmatching}.

As the neural splines are defined only in an interval, \citeauthor{durkan2019neuralsplineflows} transform values outside the interval as the identity, resulting in linear tails, by setting the boundary derivatives to $1$. 
We change the base distribution $q(\bepsilon)$ to be a truncated normal distribution defined in the interval of the splines. Therefore, all generated samples are guaranteed to stay inside the support of the splines. This is particularly useful in the early stages of optimization, when $q_{\bt}$ provides a poor approximation.

The second part of our framework focuses on the conditional distribution $q_{\bt}(\bX|\bz)$ in \rfeq{eq:cond}. We model the mean $\mu_{\bt}(\bz)$ and standard deviation $\sigma_{\bt}(\bz)$, which depend on the ``slow'' DOFs $\bz$ and must be flexible enough to capture this dependence. This density is unimodal and generally easier to learn than the multimodal density $q_{\bt}(\bz)$. To parametrize $\mu_{\bt}(\bz)$ and $\log \sigma_{\bt}(\bz)$, we employ a simple feedforward neural network, specifically a multilayer perceptron (MLP), details of which are contained in Section \ref{sec:Example}.

Lastly, the diffeomorphism $\bs{f}_{\bphi}$ in \rfeq{eq:diffeo} is learned using the aforementioned right stochastic matrix $\bs{A}_{\bphi}$. To enforce the property in \rfeq{eq:Arow} that guarantees  equivariance to rigid-body motions,  we employ  a row-wise softmax transformation to the parameter matrix $\bphi$:
\[
(\bs{A}_{\bphi})_{ij} = \frac{\exp(\bphi_{ij})}{\sum_{k=1}^N \exp(\bphi_{ik})}
\]

We implement our models using the \textsf{flowjax} \cite{ward2023flowjax} package for continuous distributions, bijective transformations, and normalizing flows using \textsf{equinox} \cite{kidger2021equinox} and \textsf{JAX} \cite{jax2018github}. Additional details can be found in the respective  numerical illustrations in the next section.  The code will be made available upon publication at \url{https://github.com/pkmtum/energy_coarse_graining_flow}. 

\section{Numerical Illustrations}
\label{sec:Example}
The following section demonstrates the capabilities of the proposed method for three use cases. First, we consider two synthetic examples: a two-dimensional double-well potential and a multimodal Gaussian mixture model. The third problem involves  the  alanine dipeptide.

\subsection{Double-Well}
\label{sec:DoubleWell}

\begin{figure}[!t]
    \centering
		\includegraphics[width=0.6\textwidth]{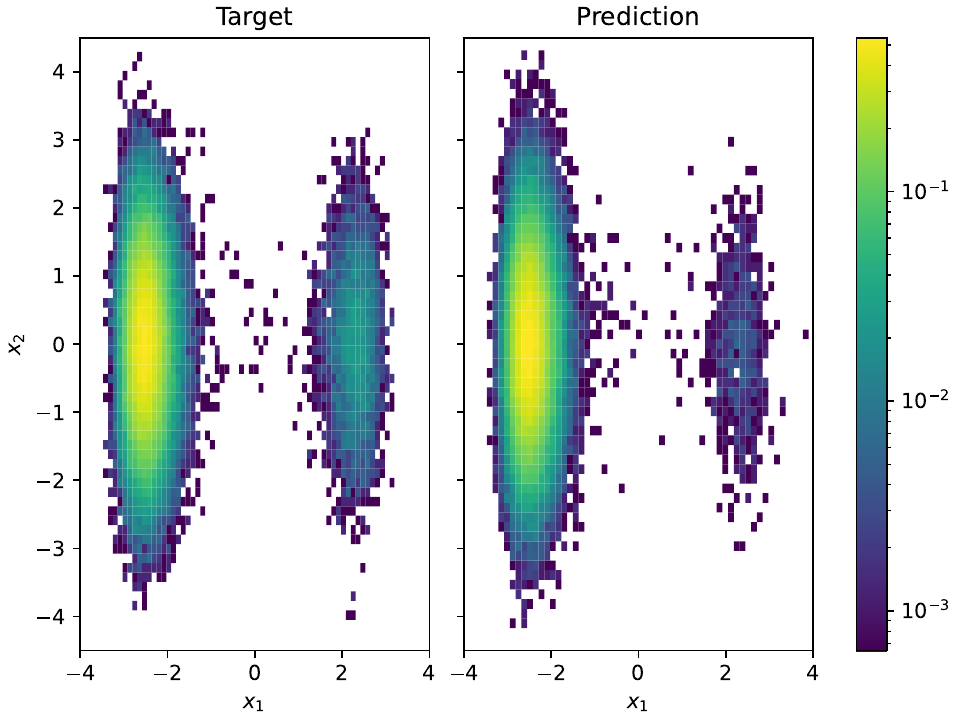}
    \caption{(Left) Histogram of all-atom samples from the target Boltzmann of \rfeq{eq:DoubleWell} and (Right) from the energy-trained approximation $q_{\bt}(\bX,\bz)$ ($\beta_{\mathrm{target}}=1$).}
    \label{fig:DWsamples}
\end{figure}

In this section, we apply our framework to a 2D double-well potential $U(\bx)$, where the two metastable states are separated by a high-energy barrier. Traditional methods, such as MD or MCMC, struggle to discover both modes in the target Boltzmann distribution since the second mode exhibits a much lower probability.
The particular form of the potential is similar to the one used by \citeauthor{noe_boltzmann_2019} and is depicted on the left side of Figure \ref{fig:DWsamples}:
\be
    U(\bx) = \frac{1}{4}x_1^4 - 3x_1^2 + x_1 + \frac{1}{2}x_2^2 \label{eq:DoubleWell}
\ee

We observe that the $x_1$-direction distinguishes between the two modes and is the slow reaction coordinate of the system. This implies that $x_1$ dictates the multimodality in the Boltzmann distribution. Therefore, we would expect that our model discovers a new set of coordinates $(z, X)$ that strongly correlate with $x_1$ and $x_2$ respectively. 
We use a $2\times 2$ right stochastic matrix $\bs{A}_{\bphi}$ to model the  transformation:
\be
 \bs{f}_{\bphi}(X,z) = \bs{A}_{\bphi} \left[ \begin{array}{c} z \\ X \end{array}\right] =
 \left[ \begin{array}{cc} 
a_1 & 1 - a_1 \\
a_2 & 1 - a_2
\end{array}\right]   \left[ \begin{array}{c} z \\ X \end{array}\right] = \left[ \begin{array}{c} x_1 \\ x_2 \end{array}\right] = \bx,
 \label{eq:DWtransf}
\ee
where $\dim(z)=\dim(X)=\dim(x_1)=\dim(x_2)=1$. In this case, we have parametrized the sought  matrix $\bs{A}_{\bphi}$ with respect to $a_1,a_2 \in [0,1]$.

We employ a normalizing flow model with the hyperparameters specified in Table \ref{tab:dw}. Since coupling layers do not operate on one-dimensional inputs, we use only the rational-quadratic spline layers without the dimensional split of the coupling layers. An MLP models the $\mu_{\bt}$ and $\sigma_{\bt}$ of the conditional distribution $q_{\bt}(X|z)=\mathcal{N}(X | \mu_{\bt}(z), \sigma^2_{\bt}(z))$ with hyperparameters in Table \ref{tab:dw_cond}. The base distribution of the flow $q(\epsilon)$ is a truncated standard normal distribution in the interval [-5, 5]. We optimize the parameters using the ADAM optimizer \cite{Kingma2015} with a learning rate $\eta_{\mathrm{SGD}}=0.001$. We train for $L=100$ update steps per tempering step with $N=500$ samples to estimate the gradients in Equations (\ref{eq:gradthetamc}) and (\ref{eq:gradphimc}) (see Algorithm \ref{alg:upd}). We found that training at the initial $\beta_0\approx 0$ for an additional $4900$ steps results in a more stable convergence of the learned map $\bs{A}_{\bphi}$ and a more successful tempering scheme.

\begin{table}[!ht]
\caption{Normalizing Flow Architecture for $g_{\bt}$ (see \rfeq{eq:repz}) in the Double-Well Potential Example.}
\label{tab:dw}
\begin{tabular}{|lllll|}
\hline 
Flow layers & MLP layers & MLP width & RQS knots & RQS Interval
\\ \hline
\hline
\rule{0pt}{3ex} $6$       & $2$       & $32$      & $8$      & $[-5, ~5]$
\\ \hline
\end{tabular}
\end{table}

\begin{table}[t]
\caption{
MLP Architecture Used for Mean and Variance of $q_{\bt}(X|z)$ (see \rfeq{eq:repX}).
}
\label{tab:dw_cond}
\begin{tabular}{|lll|}
\hline 
MLP layers & MLP width & Activation layer 
\\ \hline
\hline
\rule{0pt}{3ex} $2$ & $32$ & ReLU
\\ \hline
\end{tabular}
\end{table}

\begin{figure}[!ht]
    \centering
		\includegraphics[width=0.6\textwidth]{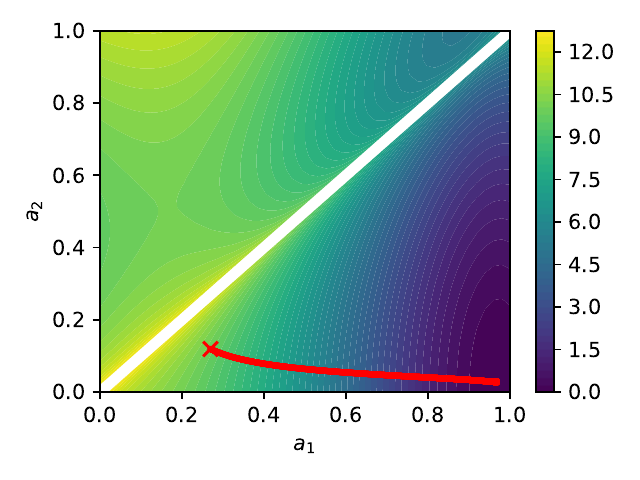}
    \caption{Contour lines of the reverse KL-divergence in \rfeq{eq:klflow} for different value pairs $(a_1,a_2)$ of the right stochastic matrix $\bs{A}_{\bphi}$ in \rfeq{eq:DWtransf}. The red $\times$ represents the starting values, and the red line represents the values during training. The final learned transformation, based on $\bs{A}_{\bphi}^{-1}$, is $z = 1.03 \cdot x_1 - 0.04 \cdot x_2$ and $X = -0.03 \cdot x_1 + 1.03 \cdot x_2$.}
    \label{fig:DWcontour}
\end{figure}

Before assessing predictive accuracy, we first discuss the learned transformation of the original DOFs (\rfeq{eq:DWtransf}).
In Figure \ref{fig:DWcontour}, we plot the KL-based learning objective of \rfeq{eq:klflow} for different values of $a_1,a_2$ (and for the optimal $q_{\bt}(X,z)$ learned). We observe the evolution of the  parameters $a_1,a_2$ from their initial values, which were randomly selected, to the minimum in the bottom right corner corresponding to $a_1=1,a_2=0$, i.e., to $z=x_1$ and $X=x_2$. The optimization was carried out with the constraint $\det(\bs{A}_{\bphi})>0$ \cite{Marzouk2016, transport_map2018}. Similar results are obtained for any other initialization point in the lower triangle, which corresponds to $\det(\bs{A}_{\bphi})>0$.

As explained in the previous sections,  energy training is highly prone to mode locking, which for this potential occurs either  at $x_1\approx -2.5$ (most often) or at $x_1\approx +2.5$. To mitigate this behavior, we use the adaptive tempering scheme   proposed (see Algorithm \ref{alg:upd}) with an inverse temperature step of $\Delta \beta_{\max}=0.05$, maximum change in KL-divergence $\delta KL_{\max}=0.1$, and target $\beta_{\mathrm{target}}=1$.

\begin{figure}
    \centering
		\includegraphics[width=.88\textwidth]{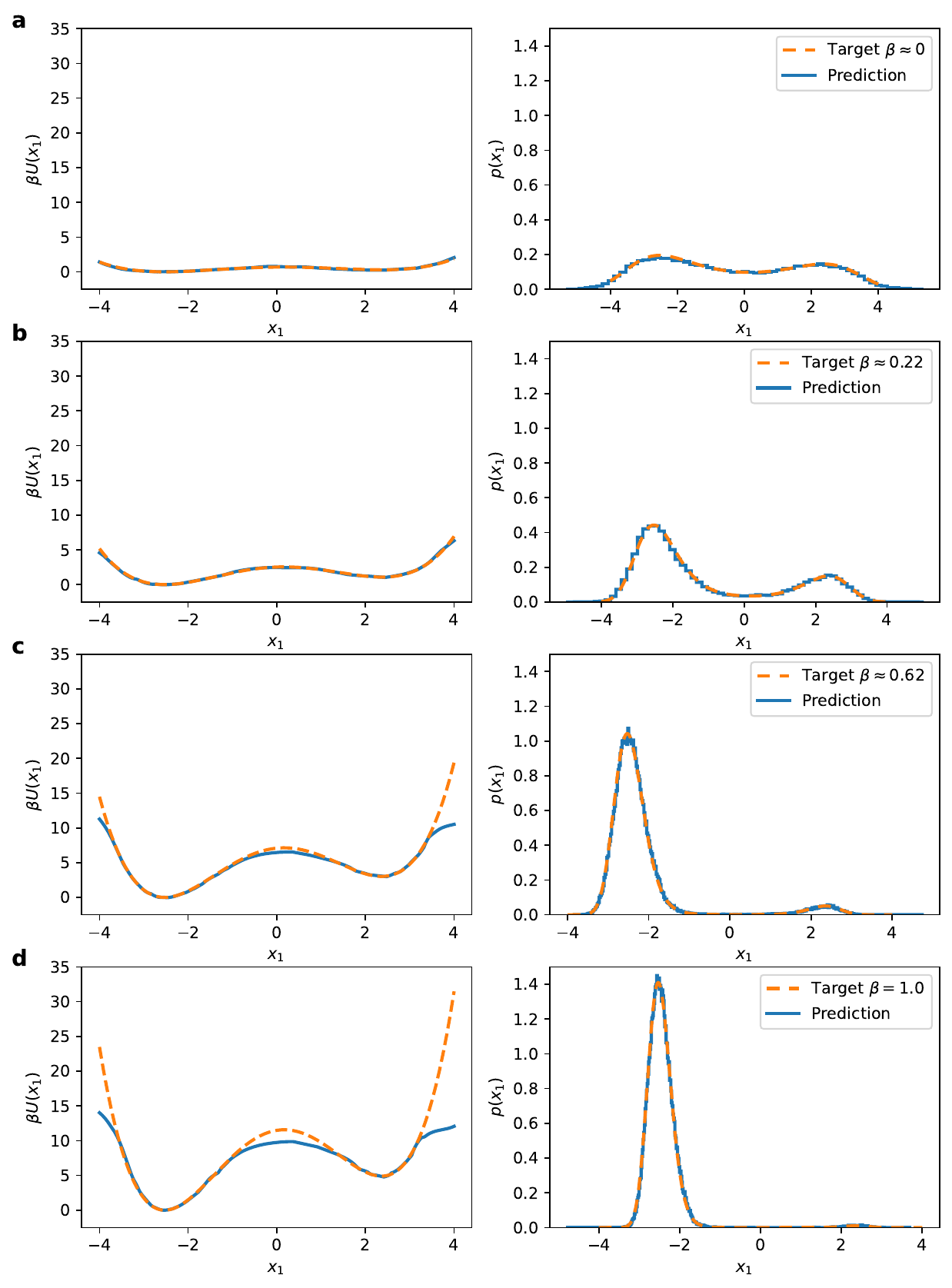}
    \caption{Left: Effective potential (PMF) $\beta U(x_1, x_2=0)=- \log p(x_1,x_2=0)$ (orange) and the predicted $U_{\bt}(x_1,x_2=0)=-\log q_{\bt}(\bs{f}_{\bphi}^{-1}(x_1,x_2=0)) + \log K_{\bphi}$ (blue) during training. Right: Histogram of samples from the marginal $p(x_1)$ (orange) and the predicted model $q_{\bt}(X,z)$ (blue). Results are shown at the inverse temperatures $(\mathbf{a}) ~\beta\approx0$, $(\mathbf{b}) ~\beta=0.2$, $(\mathbf{c}) ~\beta=0.6$, and $(\mathbf{d}) ~\beta=1$.}
    \label{fig:DW}
\end{figure}

Figure \ref{fig:DW} provides insight into the tempering scheme by displaying predictions from the model trained at various intermediate $\beta$.
The left column shows the reference potential energy $\beta U(x_1, x_2=0)=- \log p(x_1,x_2=0)$ and the predicted potential $U_{\bt}(X,z)=-\log q_{\bt}(X,z)$, which can be transformed to $U_{\bt}(x_1,x_2=0)=-\log q_{\bt}(\bs{f}_{\bphi}^{-1}(x_1,x_2=0)) + \log K_{\bphi}$ using \rfeq{eq:uphi}.
On the right column, we assess the accuracy in predicting the marginal density (and indirectly the corresponding free-energy) of the known collective variable, i.e, $x_1$.
We observe that the proposed method produces a highly  accurate approximation at the target and also at intermediate temperatures.
As one would expect, for the initial $\beta_0 = 0.01$, the target density  is close to a uniform. This accelerates the exploration of the configurational space, since the barriers between the modes are minute and easily learned by our model. As the inverse temperature $\beta$ increases, the presence of the two modes becomes more pronounced, but can be gradually captured by the flow model. At the target $\beta_{\mathrm{target}}=1$, the mode at $x_1 \approx -2.5$ has around $99\%$ of the probability mass.

Furthermore, we can readily  obtain all-atom samples  $\bx=(x_1,x_2)$ as described in Section \ref{sec:predictions}.
We obtain samples from the reference Boltzmann using a NUTS sampler \cite{nuts2014} initialized at a randomly selected location. We observe that we need $\mathcal{O}(10^8)$ energy/force evaluations in order to achieve good statistical accuracy. In Figure \ref{fig:DWsamples}, we depict two two-dimensional histograms at the target temperature $\beta_{\mathrm{target}}=1$ from the reference samples and independent samples drawn from the trained approximation $q_{\bt}(X,z)$ (see Section \ref{sec:predictions}).

We finally note that training our model required $\mathcal{O}(10^6)$ energy/force evaluations, i.e., approximately 2 of magnitude less than the reference simulations.  
With this computational cost, we obtain accurate approximations of the target Boltzmann densities at all intermediate temperatures,  
from which we can generate {\em independent, one-shot} samples, in contrast to the correlated samples obtained from the reference NUTS simulations.

\subsection{Gaussian Mixture Model}
\label{sec:GMM}

In the second synthetic example, we consider a multimodal target density that arises from a Gaussian mixture model (GMM) with three distinct modes. In particular, we partition the all-atom coordinates $\bx$  as $(\bx_{\bX},\bx_{\bz})$ and write $p(\bx_{\bX},\bx_{\bz})=p(\bx_{\bX}|\bx_{\bz}) ~p(\bx_{\bz})$ where: 
\be
p(\bx_{\bz})=\sum_{k=1}^3 w_k ~\mathcal{N}(\bx_{\bz} | \bs{m}_k, \bs{\Sigma}_k)
\label{eq:pm2}
\ee
is the mixture of three Gaussians with equal weights $w_k=1/3$, means $\bs{m}_k$ randomly sampled from a uniform distribution between $[-1, 1]^{\dim(\bz)}$, and  a diagonal covariance $\bs{\Sigma}_k=\mathrm{diag}(0.01)$. The conditional is defined as 

\be
p(\bx_{\bX}|\bx_{\bz})=\mathcal{N}(\bx_{\bX} | \bs{B} \bx_{\bz}, \bs{S}), \label{eq:GMMcond}
\ee

where the entries of the matrix $\bs{B}$ were sampled from a standard normal distribution and the covariance is diagonal $\bs{S}=\mathrm{diag}(0.01)$. We emphasize that this prepartitioning of $\bx$ is {\em not} used in the training of our model, but is solely employed for the construction of the target Boltzmann density.

In terms of dimensions, we consider two different settings: a) $\dim(\bx)=4$ with $\dim(\bx_{\bz})=2,~\dim(\bx_{\bX})=2$ and, b)  $\dim(\bx)=20$ with $\dim(\bx_{\bz})=10,~\dim(\bx_{\bX})=10$.
The lower-dimensional setting is chosen for ease of visualization, whereas the higher-dimensional setting presents substantially greater challenges for training.

\begin{table}[!ht]
\caption{Normalizing Flow Architecture for $g_{\bt}$ (see \rfeq{eq:repz}) in the GMM Example.}
\label{tab:gmm}
\begin{tabular}{|llllll|}
\hline 
Coupling layers & MLP layers & MLP width & RQS knots & RQS Interval & $\dim(\bt)$  \\ \hline
\hline
\rule{0pt}{3ex} $8$               & $2$          & $40$        & $8$         & $[-4, ~4]$  & $23272 ~(d_{\bx}=4)$ \\
 & & & & & $62600  ~(d_{\bx}=20)$  \\  \hline
\end{tabular}
\end{table}

\begin{table}[ht]
\caption{MLP Architecture Used for Mean and Variance of $q_{\bt}(\bX|\bz)$ (see \rfeq{eq:repX}).}
\label{tab:gmm_cond}
\begin{tabular}{|llll|}
\hline 
MLP layers & MLP width & Activation layer & $\dim(\bt)$ 
\\ \hline
\hline
\rule{0pt}{3ex} $2$ & $40$ & ReLU & $1924 ~(d_{\bx}=4)$ \\
& & & $2900 ~(d_{\bx}=20)$
\\ \hline
\end{tabular}
\end{table}

The details of the architecture of the normalizing flow model used for $q_{\bt}(\bz)$ can be found in Table \ref{tab:gmm} and for the conditional  $q_{\bt}(\bX|\bz)$ in Table \ref{tab:gmm_cond}. The parameters $(\bt,\bphi)$ are optimized using the ADAM optimizer with a learning rate $\eta_{\mathrm{SGD}}=0.001$. We estimate expectations of the gradients with $N=10000$ samples.

 The adaptive tempering proposed is carried out with $\delta KL_{\max}=0.1$, $\Delta \beta_{\max}=0.02$, and we use $10000$ samples to estimate the KL-divergence. We train for $L=1000$ epochs at each temperature until we reach the target $\beta_{\mathrm{target}}=1$. We note that convergence for the initial $\beta_0=0.001$ is crucial for the success of the tempering scheme. 
 Therefore, we train for an additional $19000$ update steps at this initial $\beta$.

\captionsetup[subfigure]{labelformat=empty}
\begin{figure}[!t]
    \centering
    \begin{subfigure}[b]{0.9\textwidth}
		\includegraphics[width=1.0\textwidth]{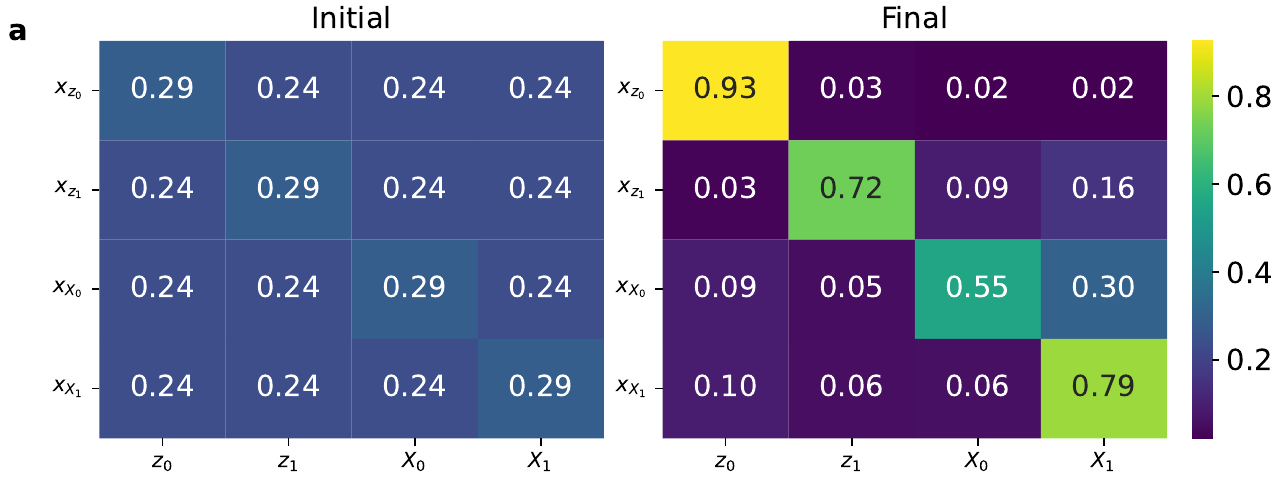}
    \end{subfigure}
    \vspace{0.5cm}
    \begin{subfigure}[b]{0.5\textwidth}
		\includegraphics[width=1.0\textwidth]{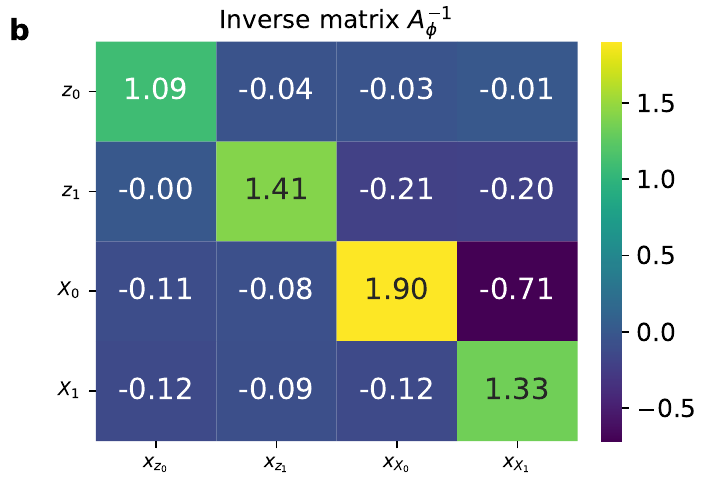}
    \end{subfigure}
    
    \caption{$(\mathbf{a})$ Right stochastic matrix $\bs{A}_{\bphi}$ at (left) initialization and (right) at target inverse temperature $\beta=1$. $(\mathbf{b})$ Inverse matrix $\bs{A}_{\bphi}^{-1}$ at $\beta=1$.}
    \label{fig:GMMA}
\end{figure}

For the lower-dimensional case ($\dim(\bx)=4)$), we show in Figure \ref{fig:GMMA} the initial and learned transformation matrix $\bs{A}_{\bphi}$ as well as its inverse $\bs{A}_{\bphi}^{-1}$ (for $\dim(\bz)=2=\dim(\bx_{\bz})$). The initial, right-stochastic matrix  $\bs{A}_{\bphi}$ has as uniform entries as possible (while ensuring that $\det(\bs{A}_{\bphi})>0$), which is achieved by adding a small positive number to the diagonal terms.
We note that the optimal $\bs{A}_{\bphi}$ identified is diagonally dominant, and this can be seen more clearly in its inverse. The learned ``slow'' variables $\bz$ are mostly associated with the $\bx_{\bz}$-coordinates, along which the target is multimodal by construction (see \rfeq{eq:pm2}). Similarly, $\bX$ are associated primarily with $\bx_{\bX}$ (see \rfeq{eq:GMMcond}).

\begin{figure}[!ht]
    \centering
    \includegraphics[width=0.5\textwidth]{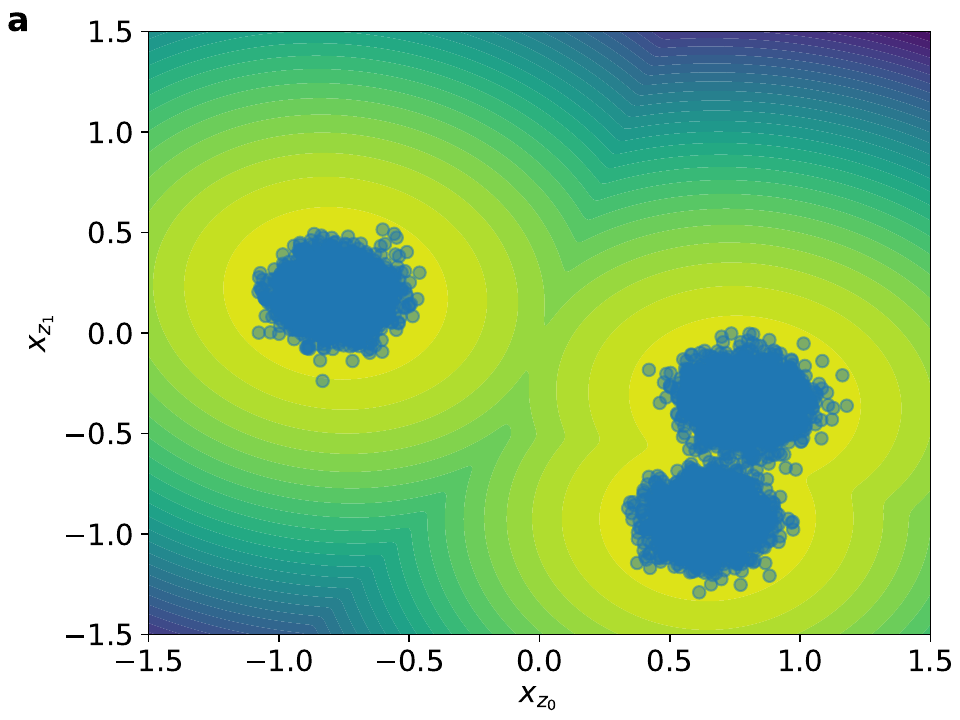}\\
    \includegraphics[width=0.8\textwidth]{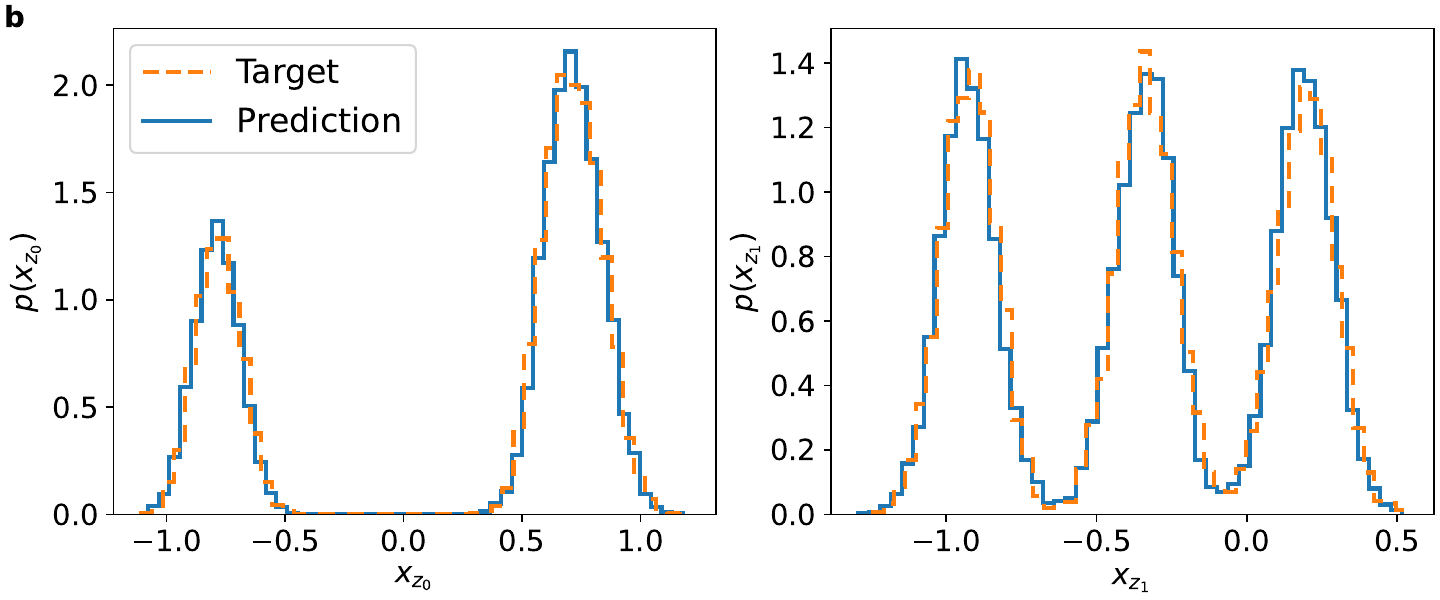}
    \caption{$(\mathbf{a})$ Scatter plot of $\bx_{\bz}$-coordinates of independent samples drawn from the learned  $q_{\bt}$. The contour lines correspond to the target, multimodal distribution $p(\bx_{\bz})$. $(\mathbf{b})$ One-dimensional marginals  of the $\bx_{\bz}$-coordinates ($\dim(\bx_{\bz})=2$) as computed from the learned $q_{\bt}$  (blue) vs target $p(\bx_{\bz})$ (orange) at  $\beta_{\mathrm{target}}=1$.}
    \label{fig:gmm2z}
\end{figure}

In Figure \ref{fig:gmm2z}, the $\bx_{\bz}$-coordinates corresponding to independent samples drawn from the trained model, i.e., $q_{\bt}(\bX,\bz)$ and transformation $\bs{f}_{\bphi}$, are depicted against contour lines from the target distribution $p(\bx_{\bz})$ above. Underneath, we compare the marginals of the $\bx_{\bz}$-coordinates. We observe that our model can accurately jointly and marginally capture all three modes of the GMM (see \rfeq{eq:pm2}).

\begin{figure}[!t]
    \centering
    \begin{subfigure}[b]{0.9\textwidth}
		\includegraphics[width=1.0\textwidth]{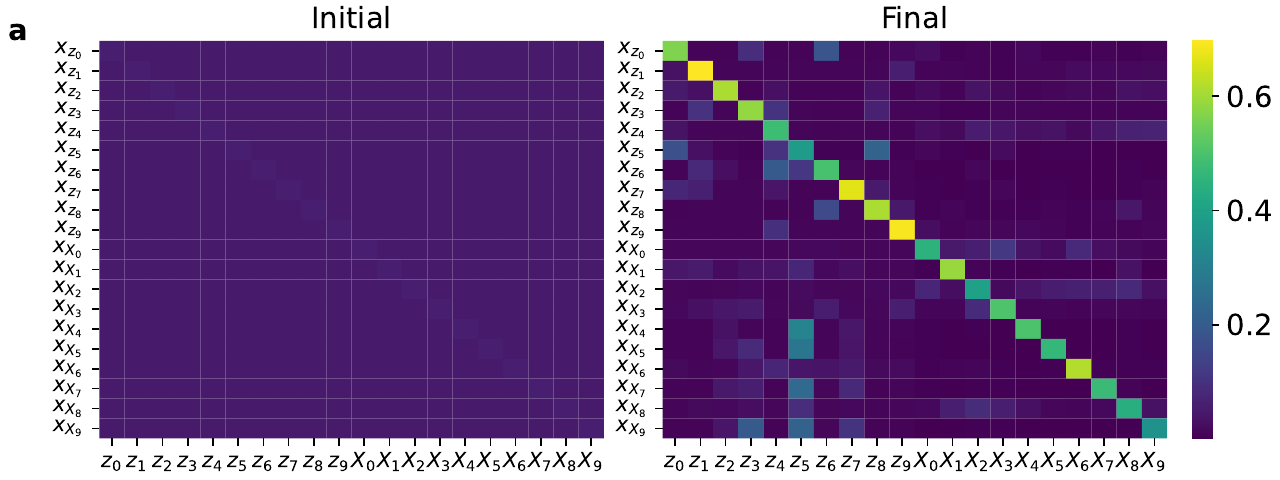}
    \end{subfigure}
    \vspace{0.5cm}
    \begin{subfigure}[b]{0.45\textwidth}
		\includegraphics[width=1.0\textwidth]{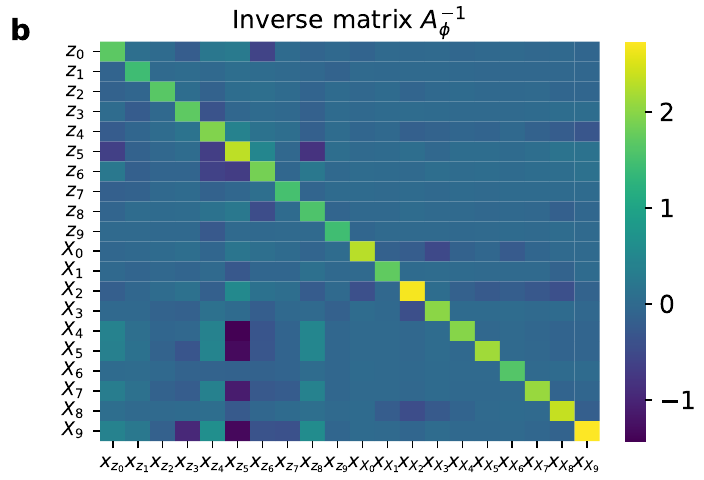}
    \end{subfigure}
    \caption{$(\mathbf{a})$ Right stochastic matrix $\bs{A}_{\bphi}$ of the $20$ dimensional $\bx$-coordinates at (left) initialization and (right) at target inverse temperature $\beta=1$. $(\mathbf{b})$ Inverse matrix $\bs{A}_{\bphi}^{-1}$ at $\beta=1$.}
    \label{fig:GMMA10}
\end{figure}

In the higher-dimensional case ($\dim(\bx)=20$), it becomes even more important to correctly identify the lower-dimensional subspace where the multimodality is concentrated. As has been reported in other works, we have also  found that the  mode-seeking behavior of the reverse KL-objective is more pronounced in this higher-dimensional setting.
The learned transformation, i.e., the right-stochastic matrix $\bs{A}_{\bphi}$ and its inverse, can be seen in Figure \ref{fig:GMMA10}. We observe that despite initializing with an almost uniform matrix, it converges to one where the new coordinates $\bz$ are mostly associated with $\bx_{\bz}$, i.e., the original coordinates along which the multimodality is concentrated. This is more obvious in the inverse $\bs{A}_{\bphi}^{-1}$, which exhibits a largely diagonal structure along the aforementioned partitioning.

 \begin{figure}
     \centering
     \includegraphics[width=1.0\textwidth]{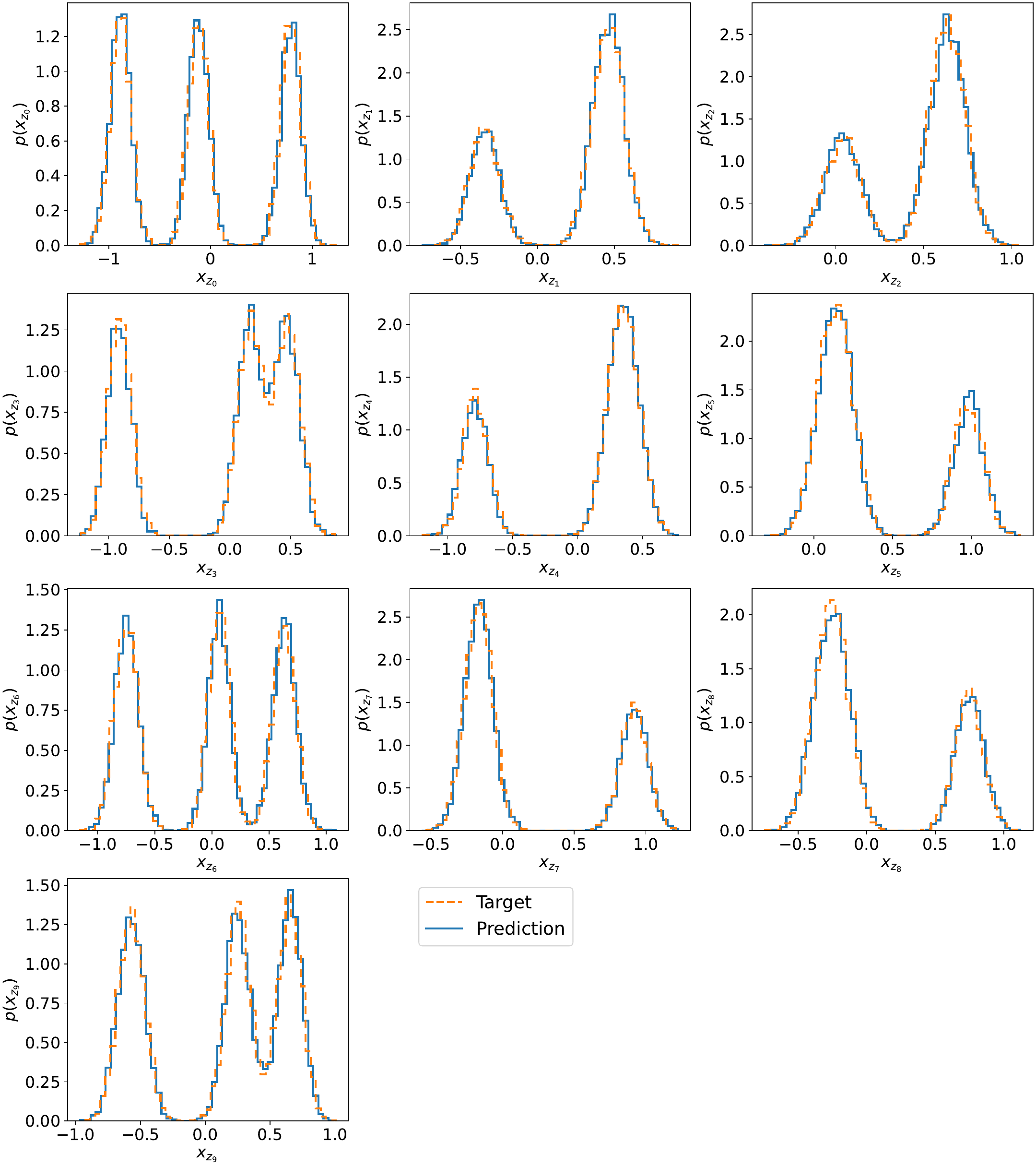}
    \caption{One-dimensional marginals of the $\bx_{\bz}$-coordinates ($\dim(\bx_{\bz})=10$) as computed from the learned $q_{\bt}$  (blue) vs target $p(\bx_{\bz})$ (orange) at  $\beta_{\mathrm{target}}=1$.}
    \label{fig:GMMmarginals10z}
\end{figure}

\begin{figure}[!ht]
    \centering
	\includegraphics[width=0.65\textwidth]{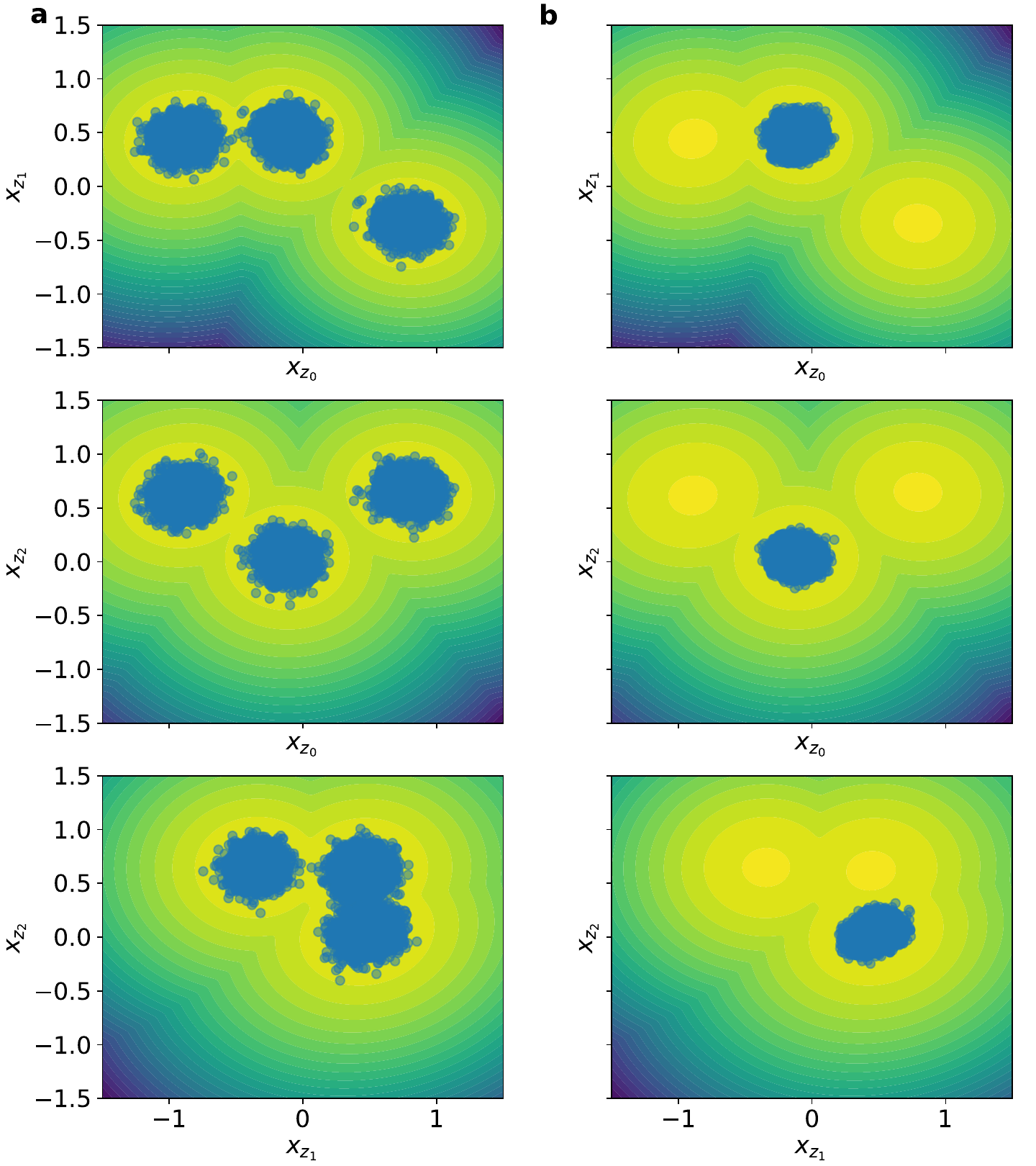}
    \caption{
    Scatter plots of different pairs of the  $10$-dimensional coordinates $\bx_{\bz}$ drawn from $q_{\bt}(\bz)$. The contour lines correspond to the target multimodal distribution $p(\bx_{\bz})$ ($\beta_{\mathrm{target}}=1$). $(\mathbf{a})$ With tempering, and $(\mathbf{b})$ without tempering. One observes that the latter leads to several modes of the target not being represented in $q_{\bt}(\bz)$. }
    \label{fig:GMMcontour10zcompare}
\end{figure}

In Figure \ref{fig:GMMcontour10zcompare}, some pairs of $\bx_{\bz}$-coordinates corresponding to independent samples drawn from the trained model $q_{\bt}(\bX,\bz)$ and transformation $\bs{f}_{\bphi}$ are depicted against contour lines from the corresponding target marginal $p(\bx_{\bz})$. On the left side, we show samples obtained after the adaptive tempering scheme proposed has been employed, and on the right side without ($\beta_{\mathrm{target}}=1$).  
We observe that the samples obtained without tempering are concentrated on one of the three modes, and the model fails to identify the other two. The application of the adaptive tempering scheme overcomes this mode-seeking behavior.  
In Figure \ref{fig:GMMmarginals10z}, we compare the learned (with adaptive tempering) one-dimensional marginals for each of the ten  $\bx_{\bz}$-coordinates against the reference ones.  We observe that  the proposed model is capable of accurately capturing all modes along all ten dimensions.

\newpage
\subsection{Alanine Dipeptide}
\label{sec:Ala}

\begin{figure}[ht]
    \centering
    \setlength{\tabcolsep}{0pt}
    \begin{tabular}{ll}
		\includegraphics[width=0.42\textwidth]{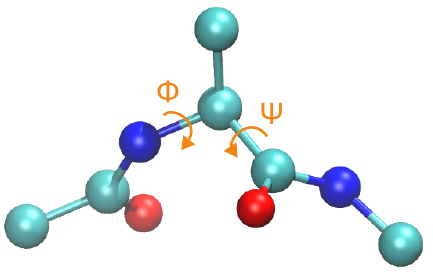}&\includegraphics[width=0.6\textwidth]{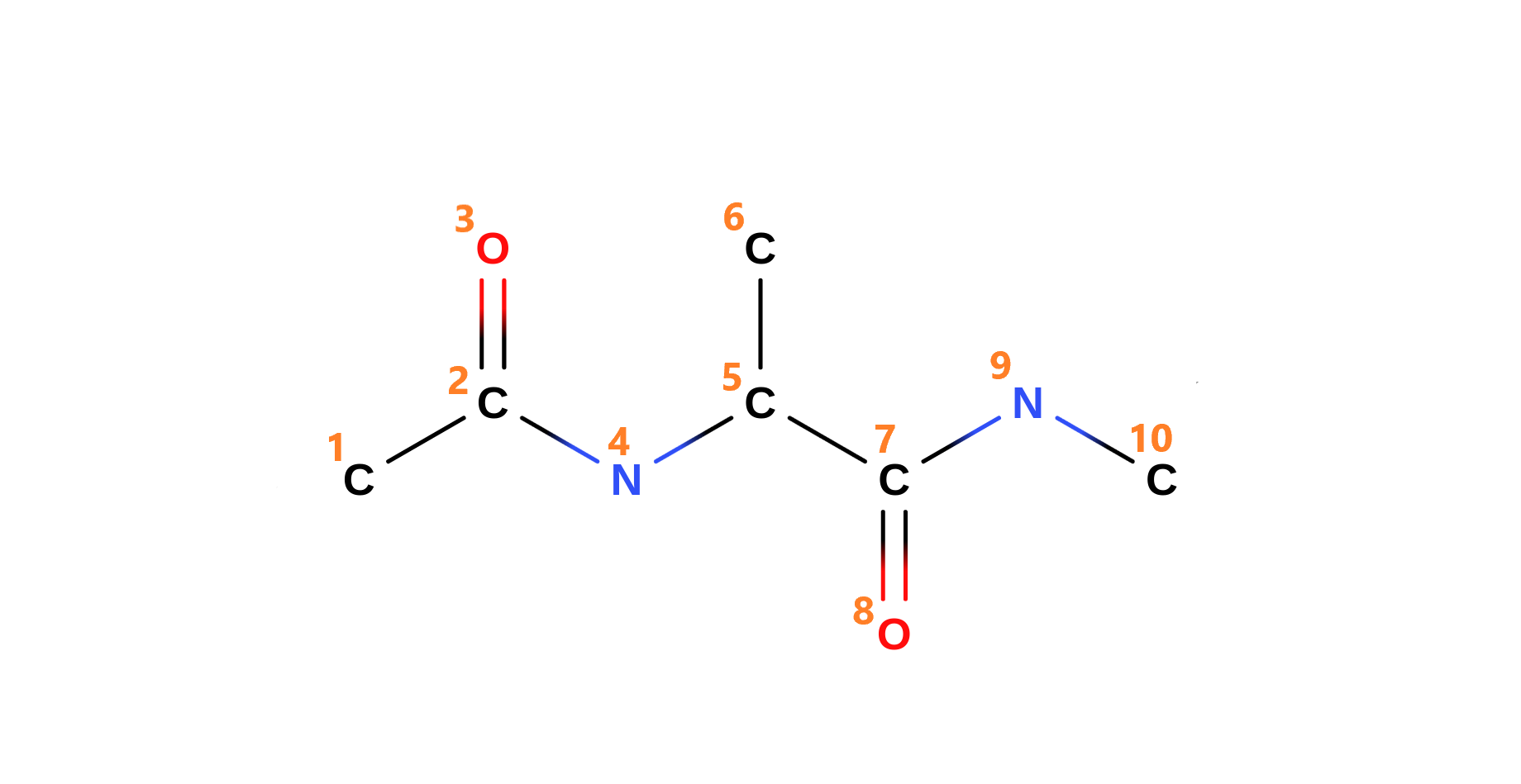}
	\end{tabular}
    \caption{Dihedral angles $(\Phi,\Psi)$ for coarse-grained alanine dipeptide (left) and atom numbering in the CG alanine dipeptide (right). We fix atom $4$ at the origin $(0,0,0)$, atom $5$ along the 3rd-axis at $(0, 0, x_3^{(5)})$, and atom $7$ on the 1–3-plane at $(x_1^{(7)}, 0, x_1^{(7)})$ in order to remove rigid body motions.}
    \label{fig:ala}
\end{figure}

The following section focuses on the data-free coarse-graining of the alanine dipeptide in an implicit solvent, a standard benchmark system with well-characterized collective variables: the dihedral angles $(\Phi, \Psi)$ (Figure \ref{fig:ala}).
As the reference, all-atom configuration $\bx$, against which all subsequent comparisons are performed, we choose an already coarse-grained version of the alanine dipeptide, where the hydrogen atoms have been removed. The employed potential energy function $U(\bx)$ is represented by the Graph Neural Network DimeNet \cite{gasteiger_directional_2022} which was trained with the relative entropy method \cite{thaler_deep_2022}. 

The reference molecule consists of $10$ atoms. We remove rigid body motions by fixing $6$ of the total $30$ Cartesian coordinates, resulting in $\dim(\bx)=30-6=24$. In particular, we fix atom $4$ at the origin $(0,0,0)$, atom $5$ along the 3rd-axis at $(0, 0, x_3^{(5)})$, and atom $7$ on the 1–3-plane at $(x_1^{(7)}, 0, x_1^{(7)})$ as shown in Figure \ref{fig:ala} \cite{schoebel_embedded_2020}. This has proven effective, particularly given the wide range of temperatures at which we train. Additionally, we fix the coordinate $x_3^{(5)}$ of the nitrogen atom $5$ to always be on the negative side in order to suppress mirror images that arise by simply flipping the direction of the bond. 
We note that the aforementioned potential cannot differentiate between the two mirror images found in alanine (\textsc{l}-form and \textsc{d}-form). Since we exclusively see the \textsc{l}-form in nature \cite{fischer_ueber_1891, blackmond_origin_2019},  we add a harmonic term \cite{wang_coarse-graining_2019}  to penalize configurations corresponding to the \textsc{d}-form.
 
The target temperature is $T=330~\mathrm{K}$ (corresponding to $\beta_{\mathrm{target}}=1$). We generated reference data by running $24$ parallel chains using the NUTS sampler, resulting in a total of $2.4 \times 10^6$ steps. To improve statistical accuracy and expedite convergence, the chains were initialized at carefully selected configurations located within free-energy wells defined by dihedral angles $\Phi$ and $\Psi$. Random initialization would require prohibitively long simulations to populate all relevant configurational regions with the correct statistical weight, even for a small molecule such as alanine dipeptide. Despite this guided initialization, the overall cost of generating the reference data was approximately $\mathcal{O}(10^9)$ energy evaluations.

In contrast, the training of the proposed method does \emph{not} rely on such initialization strategies. Our model requires approximately $\mathcal{O}(10^7)$ energy evaluations per tempering step, and about $\mathcal{O}(10^9)$ in total. While the total computational cost in terms of energy evaluations is comparable to that of generating the reference data, we emphasize two key advantages: (i) our training does not exploit preselected starting configurations, and (ii) the trained model produces a fully predictive generative representation capable of generating independent, one-shot equilibrium samples across all temperature steps.

\begin{figure}[!t]
\centering
    \begin{tabular}{ll}
        \includegraphics[width=0.45\textwidth]{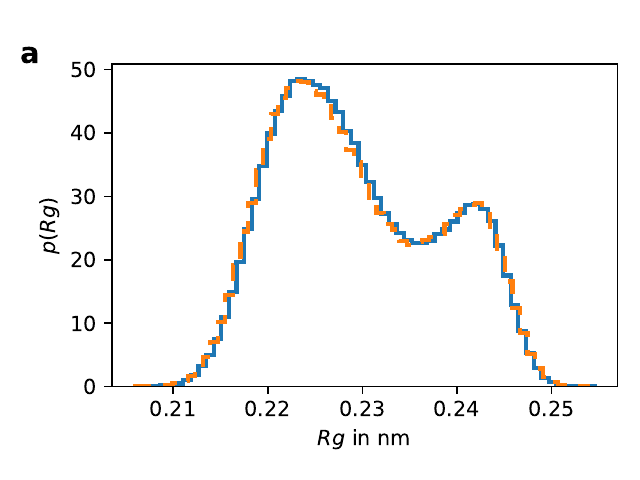}&\includegraphics[width=0.45\textwidth]{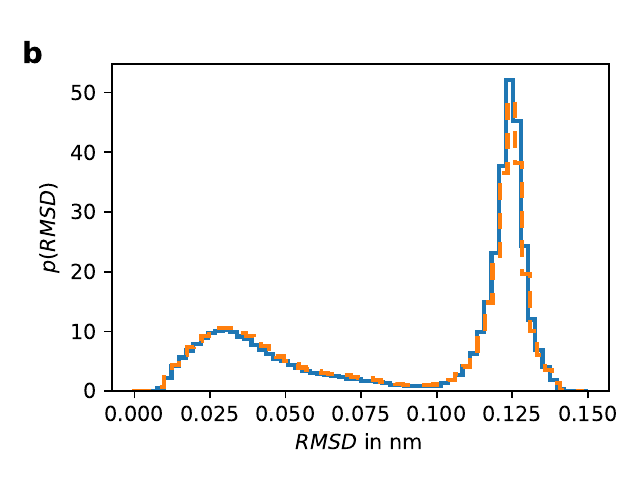}\\
        \includegraphics[width=0.45\textwidth]{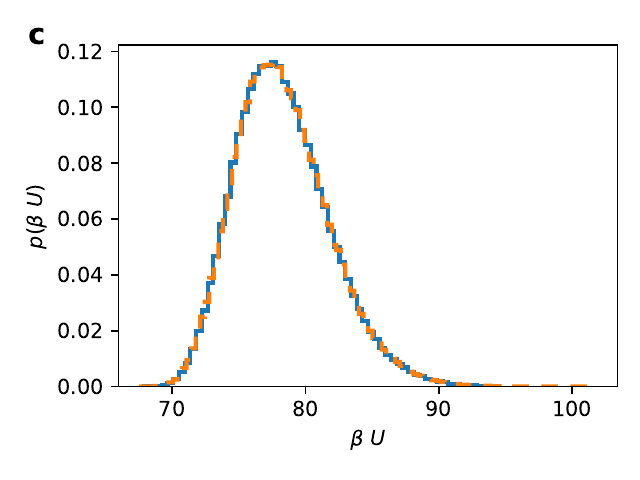}&\includegraphics[width=0.45\textwidth]{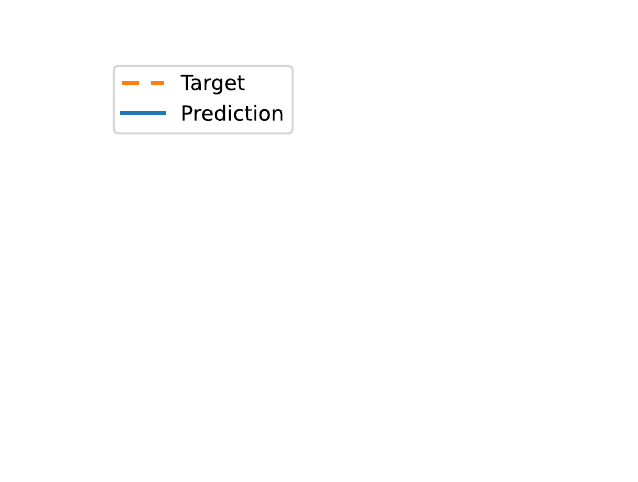}
    \end{tabular}
    \caption{Comparison of histograms for the following observables obtained from simulations of the all-atom Boltzmann density (orange) and the proposed method (blue). $(\mathbf{a})$ Radius of gyration. $(\mathbf{b})$ Root-mean-squared deviation. $(\mathbf{c})$ Energy at the target temperature.}
    \label{fig:obserables}
\end{figure}

The proposed formulation utilizes a  normalizing flow model for $g_{\bt}(\bepsilon)$, as described in Table \ref{tab:ala}. The parameters for the conditional distribution $q_{\bt}(\bX|\bz)$ can be found in Table \ref{tab:ala_cond}.
We emphasize that the proposed model has approximately {\textbf{one order of magnitude fewer parameters} compared to similar normalizing flow models that have been previously employed for (roughly) the same alanine dipeptide molecule \cite{midgley_flow_2023}.
We  optimize the parameters $\bt$ using the ADAM optimizer with a learning rate of $\eta_{\mathrm{SGD}}=5.0 \times 10^{-4}$. We skip updates with very large gradients, and clip moderate gradients according to the scheme in \citeauthor{midgley2024se3equivariantaugmentedcoupling}. We track the gradient norm of the last $50$ updates and skip gradient steps, where the norm is $10$ times higher than the median, and clip gradients, where the gradient norm is $5$ times higher than the median. This improves training stability, especially early on, when the flow model provides a very poor approximation of the target.

\begin{table}[t]
\caption{Normalizing Flow Architecture for $g_{\bt}$ (see \rfeq{eq:repz}) in the Alanine Dipeptide Example.}
\label{tab:ala}
\begin{tabular}{|llllll|}
\hline 
Coupling layers & MLP layers & MLP width & RQS knots & RQS Interval & $\dim(\bt)$ \\ \hline
\hline
\rule{0pt}{3ex} $8$               & $4$          & $64$        & $8$         & $[-4, ~4]$ & $224576$  \\ \hline
\end{tabular}
\end{table}

\begin{table}[t]
\caption{MLP Architecture Used for Mean and Variance of $q_{\bt}(\bX|\bz)$  (see \rfeq{eq:repX}).}
\label{tab:ala_cond}
\begin{tabular}{|llll|}
\hline 
MLP layers & MLP width & Activation layer & $\dim(\bt)$ 
\\ \hline
\hline
\rule{0pt}{3ex} $8$ & $90$ & ReLU & $60408$
\\ \hline
\end{tabular}
\end{table}

Convergence at the initial $\beta_0=0.0001$ is crucial and for this reason we employ $L=15000$ update steps and take $N=10000$ samples to estimate the gradients. For the adaptive tempering, we use  $\delta KL_{\max}=0.1$, $\Delta\beta_{\max}=0.005$ (see Algorithm \ref{alg:upd}).
Once the flow is trained at the initial temperature, we train for $L=1000$ update steps per tempering step until we reach the target $\beta_{\mathrm{target}}=1$.

\begin{figure}[!t]
\centering
    \begin{tabular}{ll}
        \includegraphics[width=0.85\textwidth]{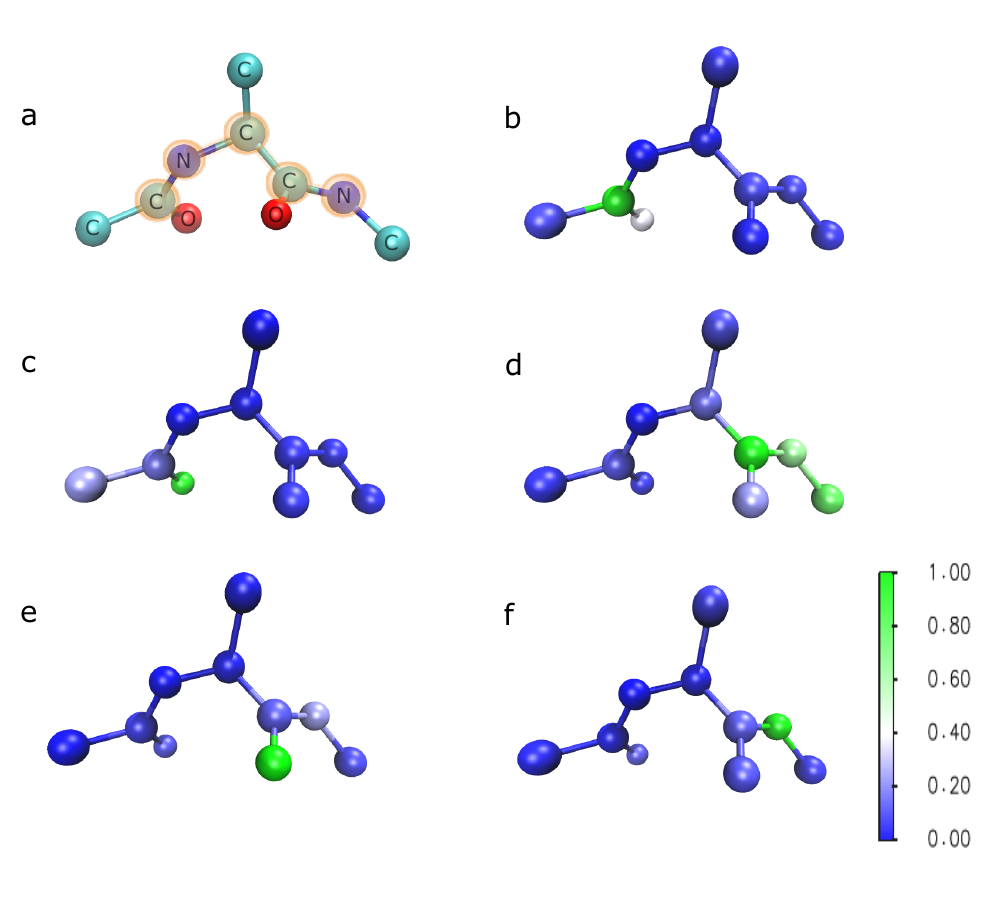}
    \end{tabular}
    \caption{Visualization of associations of pseudoatoms $\bz_{(j)}$ with actual atoms $\bx_{(i)}$ based on the learned inverse of $\bs{A}_{\bphi}^{-1}$. 
    The first configuration $(\mathbf{a})$ is selected from the reference simulation for comparison. The backbone atoms, responsible for the dihedral angles $(\Phi,\Psi)$, are highlighted in orange. The second configuration is sampled from $q_{\bt}$ and colored based on $\bs{A}_{\bphi}^{-1}$ for pseudoatom $(\mathbf{b})$  $\bz_{(0)}$, $(\mathbf{c})$  $\bz_{(1)}$, $(\mathbf{d})$  $\bz_{(2)}$, $(\mathbf{e})$  $\bz_{(3)}$, and $(\mathbf{f})$  $\bz_{(4)}$. One observes that the $\bz$-coordinates are associated with the backbone atoms and oxygen atoms of the dipeptide. }
    \label{fig:ala_pseudo}
\end{figure}

\captionsetup[subfigure]{labelformat=empty}
\begin{figure}[htbp]
    \centering
    \begin{subfigure}[b]{0.9\textwidth}
        \includegraphics[width=\textwidth]{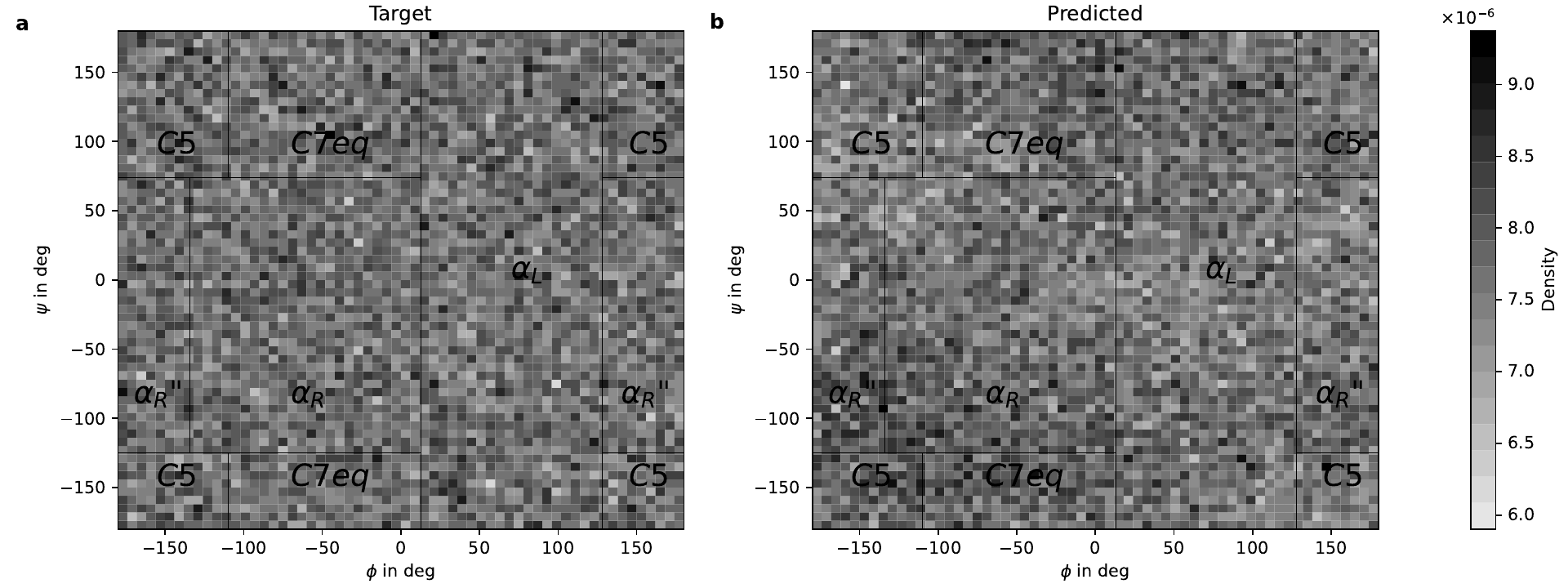}
        \label{fig:ala1}
    \end{subfigure}

    \vspace{0.5cm}

    \begin{subfigure}[b]{0.9\textwidth}
        \includegraphics[width=\textwidth]{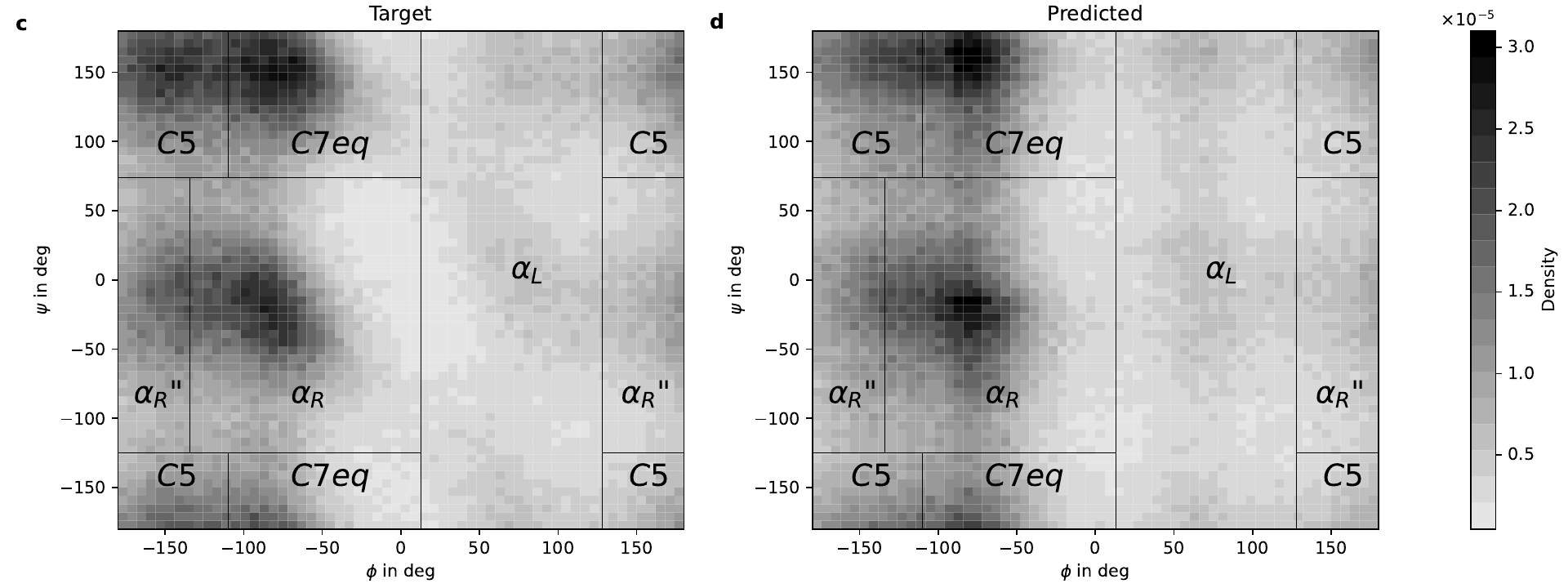}
        \label{fig:ala2}
    \end{subfigure}

     \vspace{0.5cm}

    \begin{subfigure}[b]{0.9\textwidth}
        \includegraphics[width=\textwidth]{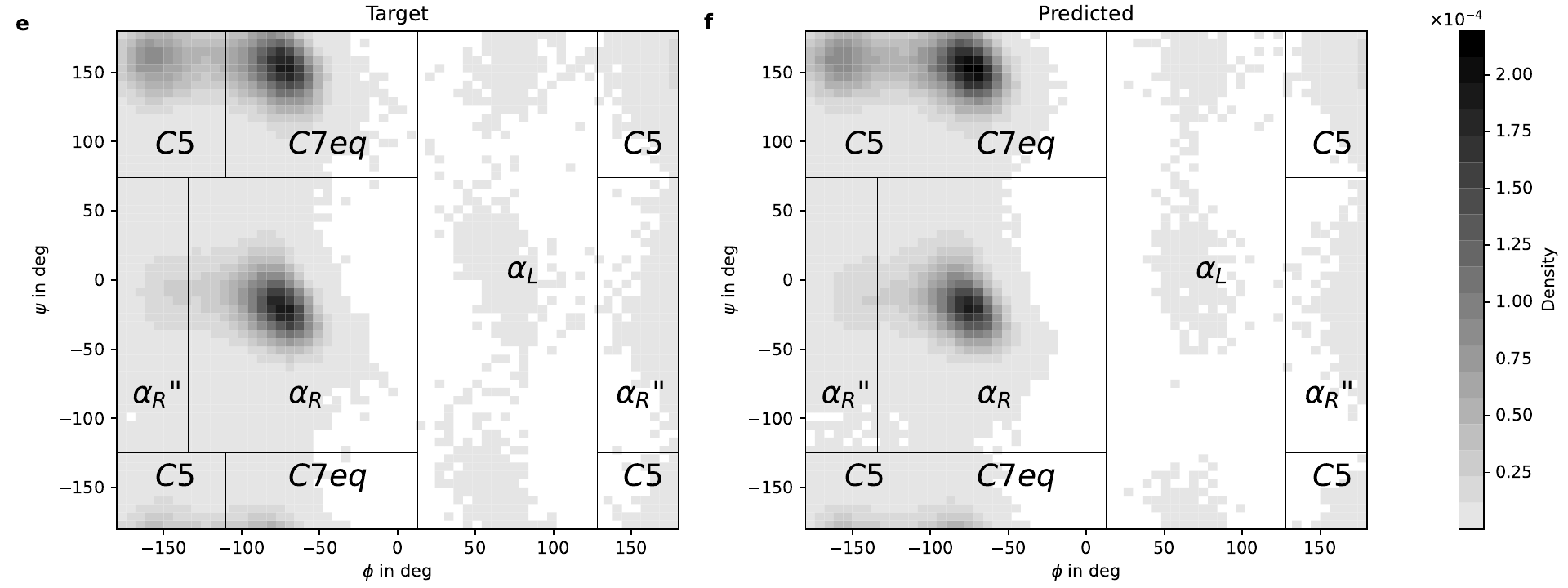}
        \label{fig:ala3}
    \end{subfigure}
    
    \caption{Ramachandran plots of the $(\Phi,\Psi)$ dihedral angle distributions obtained from the reference simulations (left) and predicted by the proposed method (right) at different inverse temperatures. Panels $(\mathbf{a}, \mathbf{b})$ correspond to $\beta \approx 0$, panels $(\mathbf{c}, \mathbf{d})$ correspond to $\beta \approx 0.22$, and panels $(\mathbf{e}, \mathbf{f})$ correspond to $\beta=1$.}
    \label{fig:rama}
\end{figure}

\begin{figure}[!t]
\centering
    \begin{tabular}{ll}
        \includegraphics[width=1.0\textwidth]{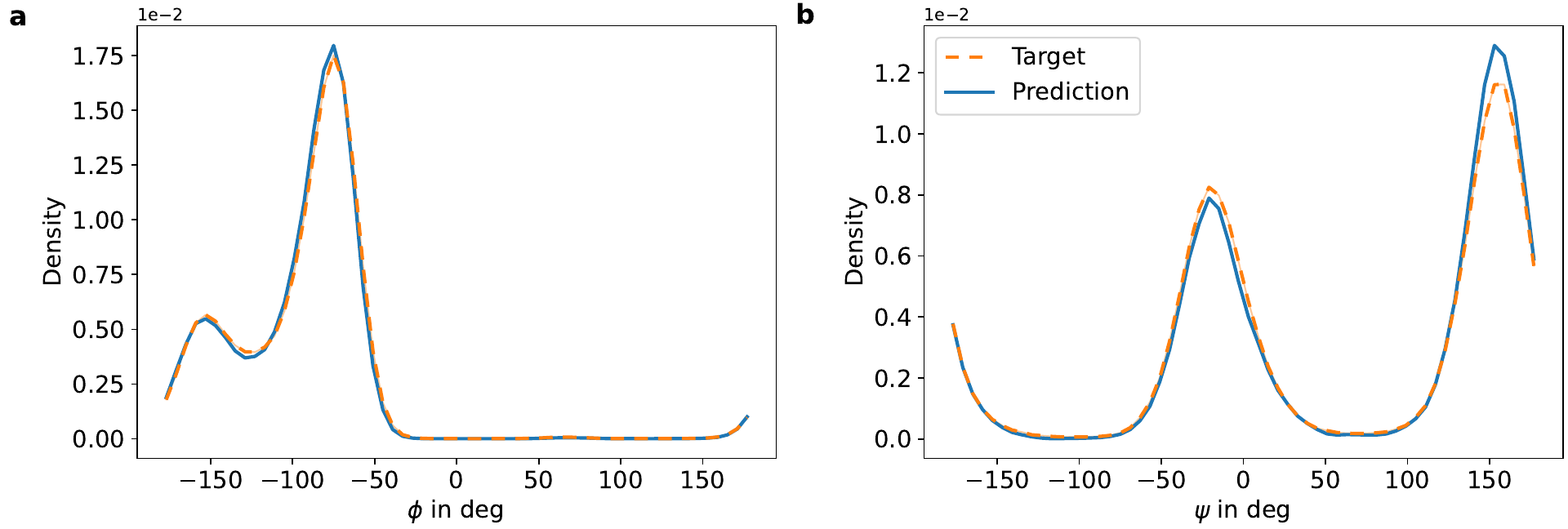}
    \end{tabular}
    \caption{Comparison of the dihedral density of $(\mathbf{a})$ $\Phi$ and $(\mathbf{b})$ $\Psi$  obtained from simulations of the all-atom Boltzmann density (orange) and the proposed method (blue).}  
    \label{fig:ala1d}
\end{figure}

In our experiments, we set $\dim(\bz)=15$ and $\dim(\bX)=9$, corresponding to $5$ and $3$ pseudoatoms, respectively. Figure~\ref{fig:ala_pseudo} visualizes the learned transformation based on $\bs{A}_{\bphi}^{-1}$. The color intensity indicates the contribution strength of each real atom $\bx_{(i)}$ to the corresponding pseudoatom $\bz_{(j)}$. We observe that the learned pseudoatoms predominantly align with the backbone atoms and also capture the contributions of the oxygen atoms. Finally, we note that the first nitrogen atom is fixed at the origin, as previously described, and is therefore excluded from the transformation.

In order to provide further insight on the identified CG coordinates $\bz$, we explore their  correlation  with the dihedral angles $(\Phi,\Psi)$ in Figure \ref{fig:latent_angle}. For each bin in the Ramachandran plot, we average the value of each individual CG-coordinate $\bz_j$ inside that bin. This gives us an indication of the expected value of $\bz_j$ given a pair $(\Phi,\Psi)$. The color indicates the average latent variable activation for all conformations in that region. We select four coordinates to highlight different correlations between coordinates and dihedral angles.

\begin{figure}[!t]
\centering
    \begin{tabular}{ll}
        \includegraphics[width=0.8\textwidth]{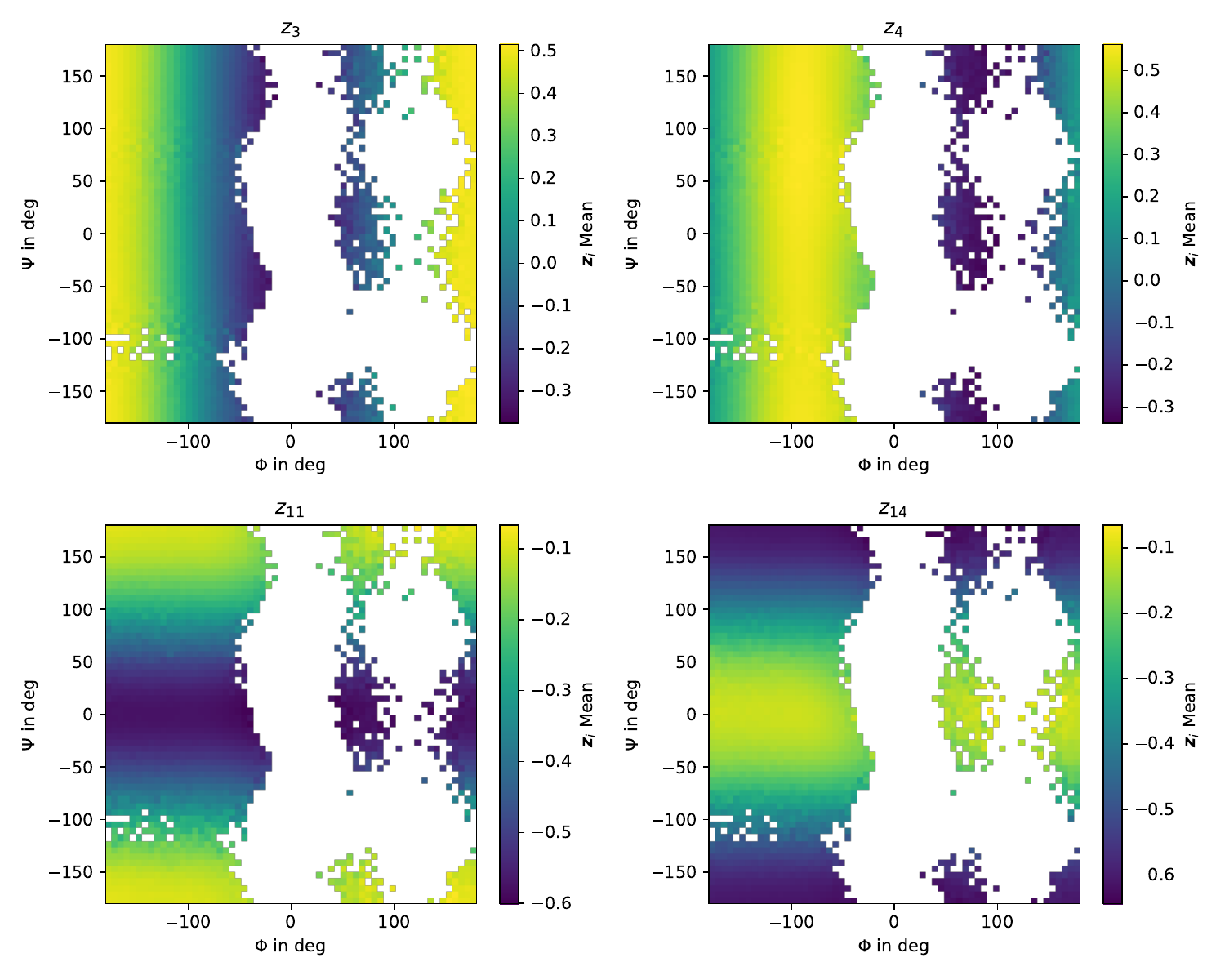}
    \end{tabular}
    \caption{Ramachandran plots of the predicted $(\Phi,\Psi)$ dihedral angle distributions. We color each bin in the histogram based on the average value of all individual CG coordinates $\bz_j$ inside that bin. Titles indicate the $\bz_j$ component. (Top left): $\bz_3$, (top right) $\bz_4$, (bottom left) $\bz_{11}$, and (bottom right) $\bz_{14}$.}
    \label{fig:latent_angle}
\end{figure}

We assess the predictive accuracy of the proposed model in terms of various observables. Firstly, results in terms of Ramachandran plots and various inverse temperatures can be seen in Figure \ref{fig:rama}. For the target temperature, we also show the one-dimensional marginals of each of the dihedral angles in Figure \ref{fig:ala1d}. We note that the proposed model approximates the density $q_{\bt}(\bz)$ of the CG coordinates $\bz$ and not of the dihedral angles depicted therein. The plots were produced using samples from the learned $q_{\bt}(\bX,\bz)$. We observe that our method is capable of finding all the relevant modes at all intermediate temperatures. We note that no pre-sampling of the target Boltzmann nor any other prior information on the location of these modes has been employed. 

We also provide histograms for the bond distances and angles of the atoms in the system in Figures \ref{fig:bond} and \ref{fig:angle}. These are in very good agreement with the reference structure.

\begin{figure}[!t]
\centering
    \begin{tabular}{ll}
        \includegraphics[width=0.8\textwidth]{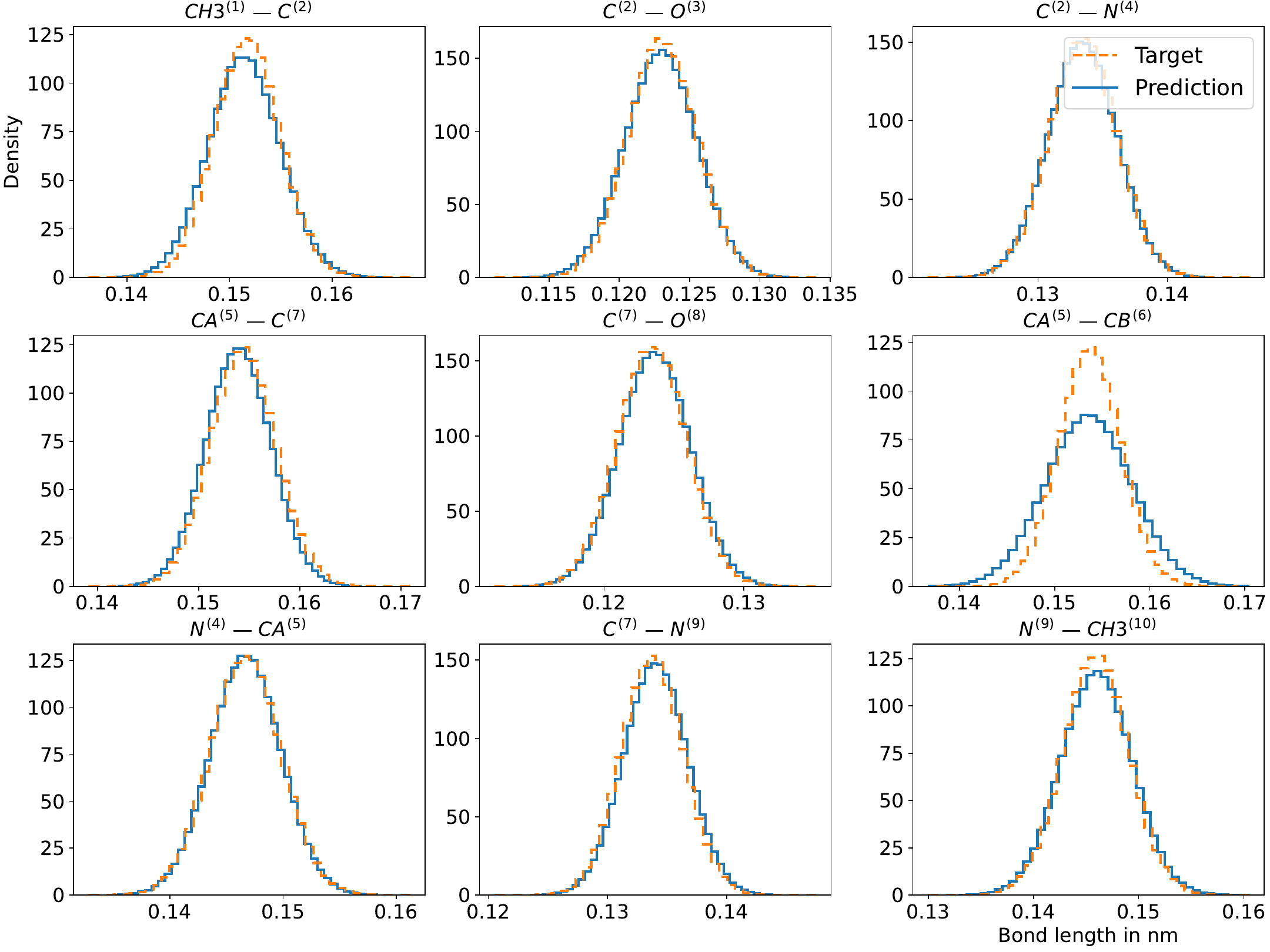}
    \end{tabular}
    \caption{Comparison of various bond-length histograms (in $nm$) obtained from simulations using the all-atom Boltzmann density (orange) and the proposed method (blue).  
    }
    \label{fig:bond}
\end{figure}

\begin{figure}[!t]
\centering
    \begin{tabular}{ll}
        \includegraphics[width=0.8\textwidth]{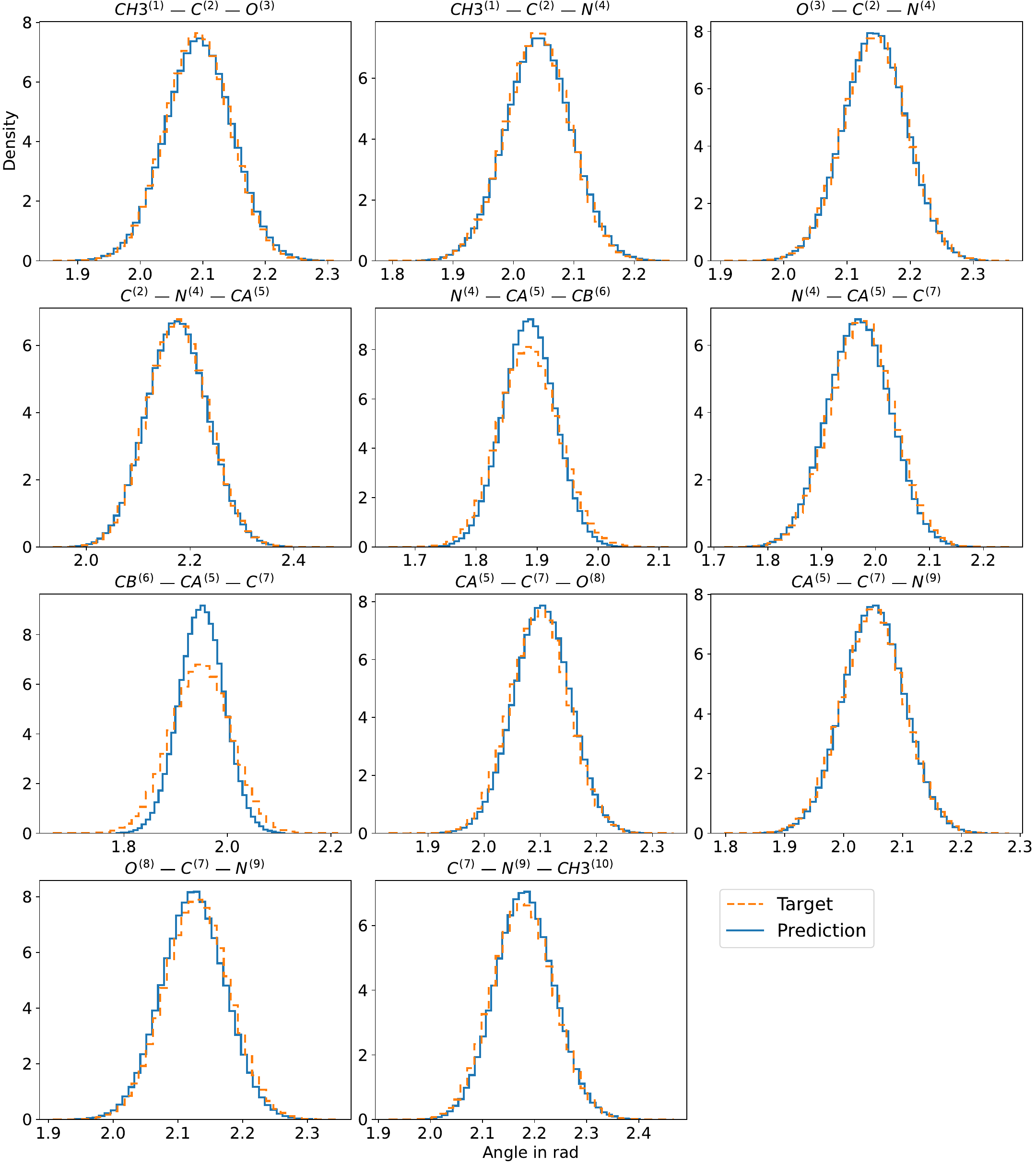}
    \end{tabular}
    \caption{Comparison of various bond-angle histograms (in $rad$) obtained from simulations using the all-atom Boltzmann density (orange) and the proposed method (blue).   
    }
    \label{fig:angle}
\end{figure}

Furthermore, we compute equilibrium distributions of three physical observables: the radius of gyration,  the root-mean-squared deviation (RMSD) \cite{fluitt_analysis_2015}, and the energy $U$ at the target inverse temperature $\beta=1$}. The radius of gyration $a_{\mathrm{Rg}}$ for a system of $N$ atoms is given by

\be
a_{\mathrm{Rg}}(\bx) = \sqrt{\frac{\sum^N_i m_i ||\boldsymbol{x}_i - \boldsymbol{x}_{\mathrm{COM}}||^2}{\sum^N_i m_i}},
\ee
where $m_i$ is the mass of each atom $i$ with Cartesian coordinates $\boldsymbol{x_i}$. The center of mass (COM) is computed as $\boldsymbol{x}_{\mathrm{COM}}=\frac{\sum^N_i m_i \boldsymbol{x}_i}{\sum^N_i m_i}$. The root-mean-squared deviation $a_{\mathrm{RMSD}}$ is calculated with respect to a reference configuration $\bx_{ref}$, which in this study was assumed to be a randomly selected position of the reference trajectory, as
\be
a_{\mathrm{RMSD}}(\bx, \bx_{ref}) = \sqrt{\frac{1}{N} \sum^N_i |\bx_i - \bx_{ref}|^2}
\ee
The histograms of these observables in Figure \ref{fig:obserables} are computed using $5000$ samples from the trained model generated as described in Section \ref{sec:predictions}.
The results for all three observables show excellent agreement with those obtained from the reference simulation.

Finally, we evaluate the quality of the generated structures by computing two scores: the bond score and the diversity score \cite{jones_diamondback_2023}. The bond score calculates how many bonded atom distances lie within $10 ~\%$ of the bond distance in the reference trajectory. The diversity score compares the average RMSD between generated structures themselves with the RMSD between generated structures and the atomistic reference \cite{wang_generative_2022, jones_diamondback_2023}. The score lies in the $[0,1]$ range. A score close to zero indicates the generated distribution is both diverse and faithful to the reference ensemble. The scores in Table \ref{tab:scores} have been obtained by averaging over $10$ trajectories of generated samples.

\begin{table}[!ht]
\caption{Bond and Diversity scores. Standard deviations are obtained by averaging the results for $10$ trajectories of generated samples.}
\label{tab:scores}
\begin{tabular}{|ll|}
\hline 
bond score ($\uparrow$) $[\%]$ & diversity score ($\downarrow$)
\\ \hline
\hline
\rule{0pt}{3ex} $0.99848 \pm 0.00001$ & $0.0106 \pm 0.0055$
\\ \hline
\end{tabular}
\end{table}

\section{Conclusions}
\label{sec:Conc}
We have presented a fully data-free, generative framework for coarse-graining that approximates the full atomistic Boltzmann distribution using only evaluations of the interatomic potential and forces. The model is trained by minimizing the reverse Kullback-Leibler divergence, and to overcome its well-known mode-seeking behavior, we introduce a novel adaptive tempering scheme based on information-theoretic criteria. Our approach {unifies the two central objectives of coarse-graining---learning a coarse-graining transformation and fitting a generative probabilistic model---into a single learning problem}, by training a bijective map from a structured latent space to full atomistic configurations. The latent space separates slow, coarse-grained variables, which capture multimodal metastable states, from fast variables representing local thermal fluctuations, enabling an accurate, one-shot reconstruction of all-atom structures. 

We demonstrated the proposed approach for benchmark problems, including a double-well potential, a Gaussian mixture model, and the alanine dipeptide molecule. Our experiments show that the method can successfully capture highly multimodal target distributions without missing modes, even in regimes where reverse KL-based training is known to fail. The adaptive tempering scheme, although it introduces computational overhead, provides an efficient and automated strategy to reduce the overall training cost.  

This framework ensures thermodynamic consistency and automatically identifies physically meaningful coarse-grained variables. By avoiding reliance on precollected MD trajectories, the method eliminates the bottleneck of costly or incomplete data generation, addressing the ``chicken-and-egg'' problem in coarse-graining.  

Although this study employed relatively lightweight neural network architectures, the framework can be extended to more expressive models, such as equivariant normalizing flows~\cite{eqflows_kohler20a, NEURIPS2021_en_eqflows} or graph neural networks~\cite{batzner_e3-equivariant_2022, geiger2022e3nneuclideanneuralnetworks}, to further improve accuracy and generalization. Overall, our results establish a scalable, interpretable, and thermodynamically faithful approach to data-free coarse-graining, providing a unified, principled solution to both sampling and back-mapping challenges in molecular modeling.

In the future, we intend to extend the proposed framework to more complex molecular systems.

In this work, we restricted our attention to the special case, where \( \bs{f}_{\bphi} \), introduced in~\rfeq{eq:diffeo}, is a learnable linear transformation. This choice provides more interpretable mappings and can be readily made equivariant with respect to rigid-body motions; several promising research directions remain open. 
Extending the analysis to nonlinear transformations (e.g., with the use of another invertible, normalizing flow model) could uncover more powerful coarse-grained representations that yield clearer separations and enable more accurate approximations.

Regardless of whether a linear or nonlinear map is used, the dimension of the CG space, $\dim(\bz)$, must still be prescribed by the user. While the objective, i.e., the KL-divergence in~\rfeq{eq:klflow}, naturally provides a score function for comparing models with different  $\dim(\bz)$, it would be highly beneficial to develop an \emph{adaptive scheme} that can automatically adjust the dimensionality, ideally starting from small values and progressively increasing $\dim(\bz)$ until no further improvements in the objective are observed. Finally, we note that although the proposed adaptive tempering scheme substantially mitigates the well-known pathologies of the reverse KL-divergence in energy-based training, it does not rule out the possibility that alternative optimization objectives (such as the Fisher or $\chi^2$ divergences discussed above) could offer a more stable and computationally efficient learning procedure.

\section*{Acknowledgement}
This work was funded by the Excellence Strategy of the Federal Government and the L\"ander in the context of the ARTEMIS Innovation Network.

\bibliographystyle{unsrtnat}
\bibliography{main}  

\begin{appendices}
\section{Computation of the information-theoretic criterion used for the adaptive tempering scheme }
\label{appendix:adap}
The following note provides a derivation of the information-theoretic criterion used for the adaptive tempering scheme proposed in Section \ref{sec:training} and associated computational details.

Based on \rfeq{eq:klflow}, we can write the numerator and denominator in \rfeq{eq:dkl} that provide $\delta KL_k$ as:
\begin{align*}
KL(q_{\bt}(\bX,\bz)|| p_{\bphi}(\bX,\bz;~\beta_{k+1})) - KL(q_{\bt}(\bX,\bz)|| p_{\bphi}(\bX,\bz;~\beta_k)) & =\Delta \beta_k \left< U_{\bphi}(\bX,\bz~;\beta) \right>_{q_{\bt}(\bX,\bz)} \\
& +\log \frac{Z_{\beta_{k+1}}}{Z_{\beta_k}} \numberthis
\end{align*}
and
\begin{align*}
KL(q_{\bt}(\bX,\bz)|| p_{\bphi}(\bX,\bz~;\beta_k))  &=  \beta_k \left< U_{\bphi}(\bX,\bz~;\beta) \right>_{q_{\bt}(\bX,\bz)}+\log Z_{\beta_k}+\left< \log q_{\bt} (\bX|\bz)\right>_{q_{\bt}(\bX,\bz)}\\
&+\left< \log ~q_{\bt}(\bz) \right>_{q_{\bt}(\bz)} \numberthis
\end{align*}
The computation of the terms excluding the partition functions, as discussed in the main text, can be carried out with Monte Carlo or analytically due to the form of $q_{\bt}$ detailed.
For the partition function $Z_{\beta_k}$
 and given the proximity of the optimized $q_{\bt}(\bX,\bz)$ to $p_{\bphi}(\bX,\bz~;\beta_k)$ we employ Importance Sampling (IS) with $q_{\bt}$ as the IS density \cite{liu_monte_2001}. In particular:
\begin{align*}
\log Z_{\beta_k} & = \log  \int e^{-\beta_k U_{\bphi}(\bX,\bz~;\beta)} ~d\bX~d\bz\\
&=  \log \int \frac{e^{-\beta_k U_{\bphi}(\bX,\bz~;\beta_k)}}{q_{\bt}(\bX,\bz)}q_{\bt}(\bX,\bz) ~d\bX~d\bz\\
&\approx \log \left( \frac{1}{N} \sum_{i=1}^N  {w}^{(i)}  \right) \numberthis
\end{align*}
where $w^{(i)}$ denote the {\em unnormalized} IS weights computed as:
\be
w^{(i)} = \frac{e^{-\beta_k U_{\bphi}(\bX^{(i)},\bz^{(i)}; ~\beta_k)} }{q_{\bt}(\bX^{(i)},\bz^{(i)})} 
\ee
where $(\bX^{(i)},\bz^{(i)})$ are i.i.d samples drawn from $q_{\bt}$. To avoid numerical underflow, we operate with the log-weights shifted by their maximum, i.e. $\tilde{w}^{(i)}=\log w^{(i)}-\log w_{max}$ where $\log w_{\max}=\max_i \log w^{(i)}$ and use the following estimator:
\be
\log Z_{\beta_k} \approx  \log \left( \frac{e^{\log w_{max}}}{N} \sum_{i=1}^N e^{\tilde{w}^{(i)}}  \right) =\log w_{max} -\log N + \log \left(\sum^N_i e^{\tilde{w}^{(i)}} \right)
\ee
Finally, with regard to the $\log \frac{Z_{\beta_{k+1}}}{Z_{\beta_k}}$ term, we have:
\begin{align*}
\log Z_{\beta_{k+1}} &= \log \int e^{-\beta_{k+1} U_{\bphi}(\bX,\bz~;\beta_{k+1})} ~d\bX~d\bz\\
&= \log \int \frac{e^{-(\beta_k + \Delta\beta_k)U_{\bphi}(\bX,\bz; ~\beta_{k+1})}}{\frac{e^{-\beta_k U_{\bphi}(\bX,\bz; ~\beta_k)}}{Z_{\beta_k}}}  \frac{e^{-\beta_k U_{\bphi}(\bX,\bz~;\beta_k)}}{Z_{\beta_k}}~d\bX~d\bz\\
&= \log Z_{\beta_k} + \log \int e^{-\Delta\beta_k U( \bs{f}_{\bphi}(\bX,\bz))} \frac{e^{-\beta_k U_{\bphi}(\bX,\bz~;\beta_k)}}{q_{\bt}(\bX,\bz)} \frac{q_{\bt}(\bX,\bz)}{Z_{\beta_k}}~d\bX~d\bz.\\
& \quad \textrm{(from \rfeq{eq:uphi} } \numberthis
\end{align*}
Hence:
\begin{align*}
\log \frac{Z_{\beta_{k+1}} }{Z_{\beta_{k}} }&= \log \int e^{-\Delta\beta_k U(\bs{f}_{\bphi}(\bX,\bz))} w(\bX,\bz) \frac{q_{\bt}(\bX,\bz)}{Z_{\beta_k}}~d\bX~d\bz\\
 & \approx \log \left( \frac{1}{N} \sum^N_{i=1} e^{-\Delta\beta_k U(\bs{f}_{\bphi}(\bX^{(i)},\bz^{(i)}))} \frac{ w^{(i)} }{Z_{\beta_k}} \right) \\
 &  = \log \left( \sum^N_{i=1} e^{-\Delta \beta_k U(\bs{f}_{\bphi}(\bX^{(i)},\bz^{(i)}))}  W^{(i)}\right) \numberthis
\end{align*}
where $W^{(i)}=\frac{w^{(i)} }{\sum_{j=1}^N w^{(j)} }$ are the {\em normalized} IS weights. As before, we can employ the shift of the log-weights to avoid numerical underflow problems in the computation of $W^{(i)}$.
\end{appendices}

\end{document}